\documentclass[fleqn,usenatbib]{mnras}

\usepackage{newtxtext,newtxmath}
 
\usepackage[T1]{fontenc}
\usepackage{ae,aecompl}

\usepackage[authoryear]{natbib}
\usepackage[T1]{fontenc}
\usepackage{ae,aecompl}
\usepackage[dvipsnames]{xcolor}
\usepackage{psfrag,color}
\usepackage[latin1]{inputenc}
\usepackage{graphicx}	
\usepackage{amsmath}	
\usepackage{amssymb}	

    \usepackage{longtable}
    \usepackage{tabu}
    \usepackage{threeparttable} 
    \usepackage{threeparttablex} 
    
    \usepackage{xcolor}

\title[The N and O gradients in the inner Galactic disc]{On the radial abundance gradients of nitrogen and oxygen in the inner Galactic disc}

\author[K. Z. Arellano-C\'ordova et al.]{
K. Z. Arellano-C\'ordova$^{1}$\thanks{E-mail: karlaz$\_$ext@iac.es and karlaz@inaoep.mx (KZAC)}
C. Esteban,$^{1,2}$
J. Garc\'ia-Rojas$^{1,2,}$
and J. E. M\'endez-Delgado$^{1,2}$
\\
$^{1}$Instituto de Astrof\'{\i}sica de Canarias, E-38200 La Laguna, Tenerife, Spain \\
$^{2}$Departamento de Astrof\'{\i}sica, Universidad de La Laguna, E-38206 La Laguna, Tenerife, Spain\\
}

\date{Accepted XXX. Received YYY; in original form ZZZ}

\pubyear{2020}

\begin{document}
\label{firstpage}
\pagerange{\pageref{firstpage}--\pageref{lastpage}}
\maketitle

\begin{abstract}
 We present optical spectra of nine Galactic \ion{H}{ii} regions observed with the 10.4 m Gran Telescopio Canarias telescope and located at Galactocentric distances ($R_G$) from 4 to 8 kpc. The distances of the objects have been revised using  \textit{Gaia} DR2 parallaxes. We determine the electron temperature for all the nebulae, which allows a precise computation of their ionic abundances. We have included published data of an additional sample of Galactic \ion{H}{ii} regions, providing a final data set of 42 objects. The shape of the radial gradients of O/H and N/H is linear and constant, discarding any substantial change of the slope, at least for $R_G$ between 4 and 17 kpc. The small dispersion of the O/H and N/H values with respect to the computed gradients implies the absence of significant azimuthal variations of the chemical abundances, at least in the quadrant covered by our observations. We find an almost flat N/O $versus$ O/H diagram relation. This result is not observed in other nearby spiral galaxy except M31. Finally, we compare our computed gradients with those obtained using far infrared (FIR) spectra. We confirm the significant offset in the N/O distribution between the optical and FIR observations. Possible explanations involve ionization correction factors and the strong dependence on density of the abundance determinations based on FIR lines.    
\end{abstract}




\section{Introduction}
\label{sec:intro}

A proper knowledge of the radial distribution of metals (elements heavier than helium) across the disc in spiral galaxies represents an important constraint for the study of the chemical enrichment of the interstellar medium (ISM). \ion{H}{ii} regions are excellent tracers of the present-day chemical content of the ISM in galaxies, and deep observations of these objects provide robust determinations of the chemical composition of the ionized gas. In particular, Galactic \ion{H}{ii} regions show a large number of emission lines $-$collisionally excited lines (CELs) and recombination lines (RLs)$-$ allowing the computation of physical conditions (electron temperature and density) and chemical abundances of different heavy elements  \citep[e.g.][]{Esteban:2004, Garcia-Rojas:2007a}. 

 Oxygen and nitrogen, the third and sixth most abundant elements in the Universe, play an important role in the study of the chemical evolution of galaxies due to the different time scales they are ejected to the ISM \citep{Garnett:1990, Vincenzo:2016,Vincenzo:2018, Berg:2020, Esteban:2020}. 
In our Galaxy, the radial distributions of N and O have been widely studied for several groups, they follow linear relations whose slopes range from $-0.040$ to $-0.060$ dex kpc$^{-1}$ for O/H, from $+0.043$ to $-0.090$ dex kpc$^{-1}$ for N/H and from $0.004$ to $-0.047$ dex kpc$^{-1}$ for N/O \citep{Shaver:1983,  Vilchez:1996, Deharveng:2000, Rudolph:2006, Fernandez-Martin:2017, Esteban:2017, Esteban:2018, Arellano-Cordova:2020a}.  
The most recent results of the O abundance gradient have been reported by \citet{Esteban:2018} and \citet{Arellano-Cordova:2020a}, who use a sample of Galactic \ion{H}{ii} regions with high-quality spectra covering a wide range in Galactocentric distances ($R_G$) between 6 and 17 kpc. The reassessments of O and N gradients of \citet{Arellano-Cordova:2020a} include new determinations of the $R_{G}$ based on \textit{Gaia} DR2 parallaxes. These authors reported an O abundance gradient of $-0.037\pm0.009$ dex kpc$^{-1}$, showing significantly lower dispersion around the fit than previous results in the literature \citep{Rudolph:2006, Fernandez-Martin:2017}, supporting the previous results of \citet{Esteban:2018}. \citet{Arellano-Cordova:2020a} also reported slopes for N/H and N/O gradients of $-0.049$ dex kpc$^{-1}$ and $-0.011$ dex kpc$^{-1}$, respectively.

The results of the studies devoted on the calculation of Galactic gradients are based on different samples of Galactic \ion{H}{ii} regions and ranges of $R_G$. Morever, the procedures to calculate physical conditions and ionic abundances are also different. In the literature, we find  different methodologies: i) electron temperatures, $T_{\rm e}$, based on radio observations and abundances derived from optical observations \citep{Shaver:1983, Deharveng:2000}; ii) a combination of radio and far-IR (FIR) observations \citep{Rudolph:2006}; iii) electron density, $n_{\rm e}$ and abundances based on FIR observations \citep[e.~g.][]{Simpson:1995, Martin-Hernandez+2002}; iv) a combination of radio and FIR observations with theoretical models \citep{Afflerbach:1997}, and v) physical conditions and abundances based on optical data \citep{Esteban:2017, Fernandez-Martin:2017, Esteban:2018, Arellano-Cordova:2020a}. Galactic \ion{H}{ii} regions are objects with large angular diameters, implying that the optical, FIR and radio observations do not necessarily cover the same zone of the nebula. Therefore, abundances derived from a multi-wavelength approach might be in disagreement with those calculated using pure optical results, increasing the dispersion of the data around the gradient (see Section~\ref{sec:abundance-gradients}). 

On the other hand, most of the optical studies of Galactic \ion{H}{ii} regions are not able to reach the innermost disc of the Milky Way due to the high dust extinction. That becomes an important issue for a better constraint of the true shape of the O abundance gradient along the disc. However, FIR  observations of \ion{H}{ii} regions provide an alternative to get information on the chemical composition of the ISM at those $R_{G}$ \citep{Peeters:2002, Martin-Hernandez+2002}. Such abundances are based on line intensity ratios of FIR fine-structure lines as  [\ion{N}{iii}], [\ion{O}{iii}], [\ion{Ne}{iii}] and [\ion{S}{iii}], with wavelengths between 12.8 and 88.7 $\mu$m. Although those lines are much less sensitive to $T_{\rm e}$ than optical CELs of the same ions \citep{Osterbrock:2006, Martin-Hernandez+2002}, the abundances calculated using FIR might be affected by other different problems. For example, there are not [\ion{O}{ii}] lines in the FIR and it is not possible to calculate the O$^{+}$/H$^{+}$ ratio, whose abundance dominates in regions with high O abundances, which are located at smaller $R_{G}$. Therefore, it is necessary the use of an ionization correction factor (ICF) scheme to correct for the missing O$^+$ abundance to determine the total O abundance from FIR observations. These ICFs can reach very high values for low-ionization conditions \citep{Martin-Hernandez+2002}, increasing the uncertainties in the total abundance estimates. Due to this, there is a clear advantage for the abundances calculated using optical data. By simply adding the contribution of the O$^+$ and O$^{++}$  abundances, we can obtain direct values of 12+log(O/H). 

Another issue regarding FIR observations is that the intensity of fine-structure lines depend strongly on the $n_{\rm e}$ of the nebula  \citep{Stasinska:2005, Osterbrock:2006}. \citet{Martin-Hernandez+2002} derived densities for a sample of compact \ion{H}{ii} regions using FIR observations. These authors noted that the densities derived by using the intensity ratio of [\ion{O}{iii}] $\lambda$$\lambda$51.8, 88.3 $\mu$m lines provided values of density much lower than those obtained from radio observations in compact regions, affecting significantly the abundances calculated. 

In the case of N abundances, both optical and FIR spectra suffer from the missing detection of one of the two ionization states of this element that are expected in \ion{H}{ii} regions. In the optical range is possible to detect [\ion{N}{II}] $\lambda\lambda$6548, 6583 \AA\, while in FIR observations the detected line is [\ion{N}{iii}] 57 $\mu$m and, in some cases, [\ion{N}{II}] 121.7 $\mu$m although is usually very faint \citep{Martin-Hernandez+2002}. Therefore, the computation of the total abundances of N requires the use of an ICF scheme. \citet{Esteban:2017} and \citet{Arellano-Cordova:2020a} observed a group of low-ionization \ion{H}{ii} regions for which no ICF is needed to derive their N/H ratio. \citet{Arellano-Cordova:2020a} analyzed  different ICF schemes for N, concluding that the classical ICF of \citet{Peimbert:1969} works well for \ion{H}{ii} Galactic regions. On the other hand, the ICF schemes used in FIR observations are not easy to evaluate and a high correction is necessary to account for the missing N$^{+}$. Since N$^{+}$ becomes dominant at smaller $R_G$, a large correction for this ion implies a more uncertain N abundance. This also implies a significant discrepancy between the N abundances calculated from both optical or FIR measurements specially at small $R_G$ values, as \citet{Rudolph:2006} pointed out. Finally, it is important to stress that although optical spectra are not free of difficulties concerning the computation of chemical abundances,  the use of high-quality spectra, updated atomic data \citep{Esteban:2017, Esteban:2018, JuandeDios:2017, Rodriguez:2020} and the reassessment of ICF schemes using photoionization models (Amayo et al. in preparation) have improved significantly the  abundance calculations of different elements, which contribute to a better constraint of the radial abundance gradients along the Galactic disc. 

In this work, we present deep high-quality optical observations of nine Galactic \ion{H}{ii} regions located at $R_G$ between 4 and 8 kpc. These observations, taken with the 10.4 m Gran Telescopio Canarias telescope, allow us to compute and characterize the chemical abundances of N and O, allowing to constrain the radial abundance gradients of these elements in the inner Galactic disc, the zone inside the Solar radius ($R_G < 8.2$ kpc). It is important to address that optical studies of \ion{H}{ii} regions are limited at $R_G > 4$ kpc, because dust extinction prevents studying the radial abundance gradient at lower Galactocentric distances from optical spectra. 

The structure of this paper is as follows. In Section~\ref{sec:observations} we describe the observations, reduction procedure and the additional sample of Galactic \ion{H}{ii} compiled from the literature. We also present a reassessment of the distances of the new objects based on {\it Gaia} parallaxes of the second data release (DR2) \citep{gaiadr2}. In Section~\ref{sec:calculations}, we present the physical conditions and ionic and total abundances. In Section~\ref{sec:abundance-gradients} we present the radial abundance gradients of N, O and N/O and the comparison of our results with those with FIR observations. Finally, in Section~\ref{sec:conclusions} we summarize our main conclusions.


\section{Observations and data reduction}
\label{sec:observations}

\begin{figure*} 
\begin{center}
    \includegraphics[width=0.75\textwidth, trim=30 0 30 0,  clip=yes]{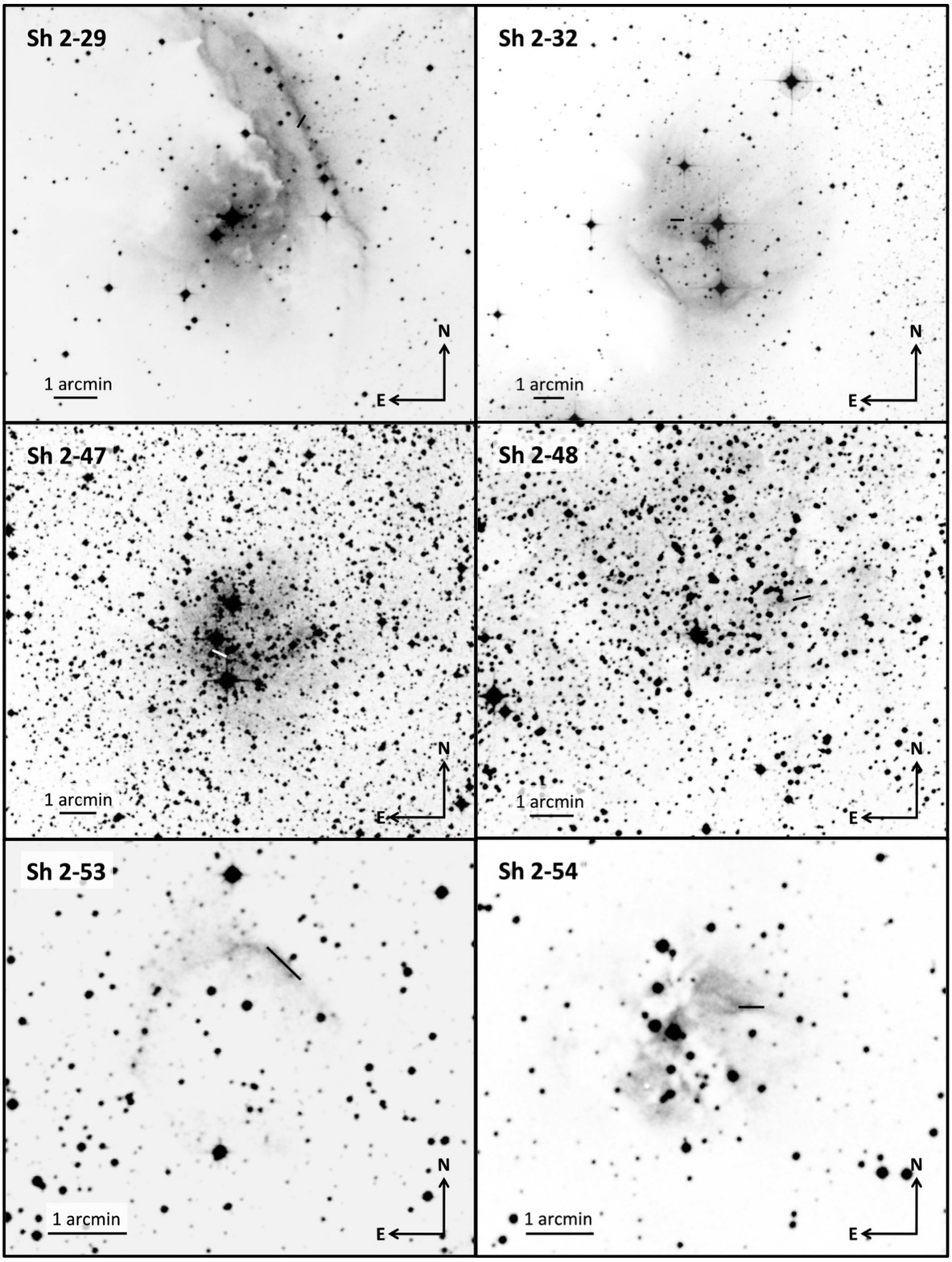}
    \caption{Images of Sh\,2-29, Sh\,2-32, Sh\,2-47, Sh\,2-48, Sh\,2-53 and Sh\,2-54. The black/white lines show the position and length of the aperture extracted for each object. The images correspond to blue (Sh\,2-54), colored (Sh\,2-29, Sh\,2-32, Sh\,2-48 and Sh\,2-53) or red (Sh\,2-47) images of the Second Digitized Sky Survey (DSS-2) taken from the Aladin interactive sky atlas \citep{Bonnarel:2000}. }
    \label{fig:maps1}
   \end{center}
    \end{figure*}

\begin{figure*} 
\begin{center}
    \includegraphics[width=0.75\textwidth, trim=30 0 30 0,  clip=yes]{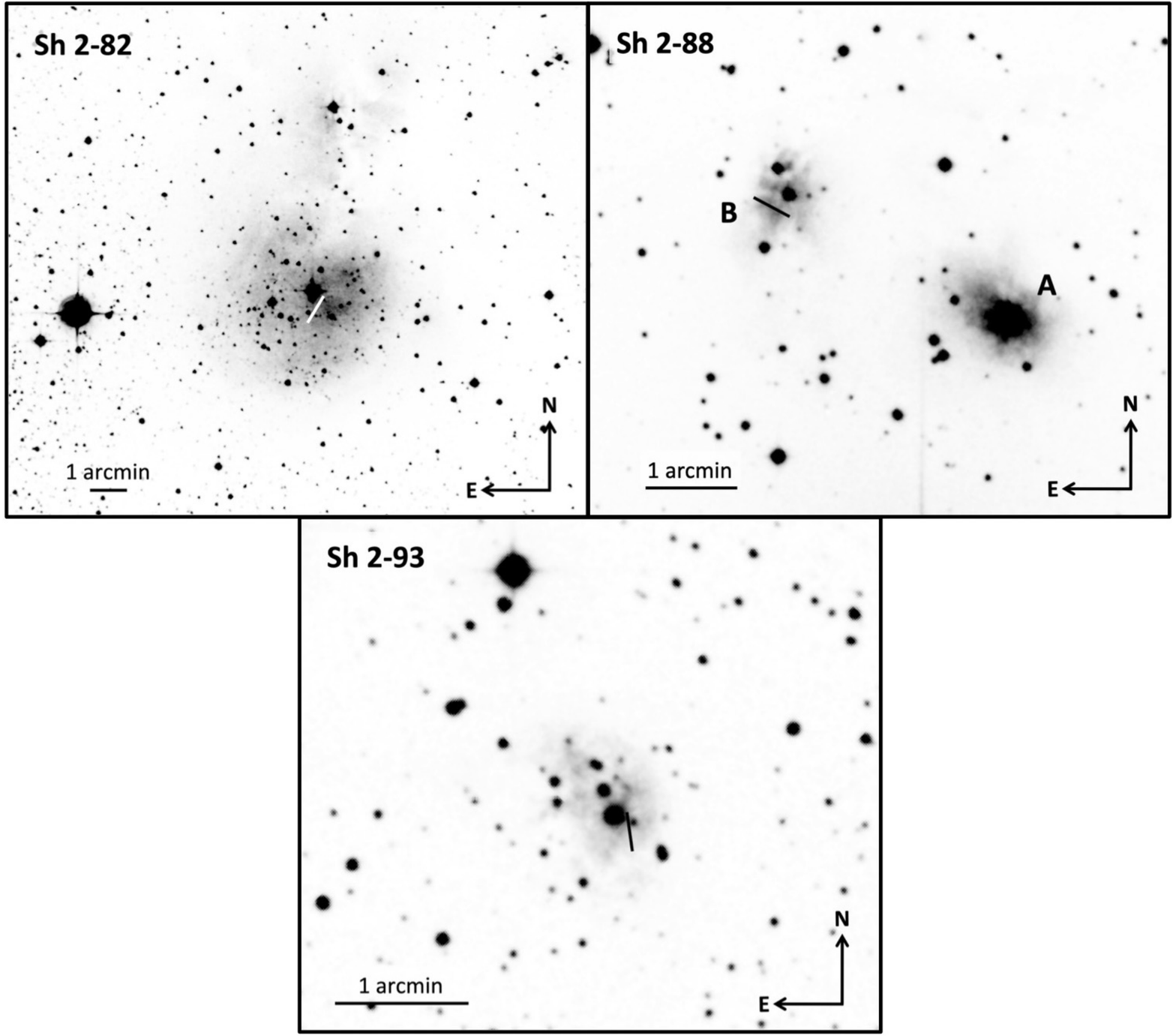}
    \caption{Images of Sh\,2-82, Sh\,2-88, and  Sh\,2-93. The black/white lines indicate the position and length of the area extracted for each object. The images correspond to blue (Sh\,2-88, Sh\,2-93) or red (Sh\,2-82) images of the Second Digitized Sky Survey (DSS-2) taken from the Aladin interactive sky atlas \citep{Bonnarel:2000}. }
    \label{fig:maps2}
   \end{center}
    \end{figure*}

The sample objects were observed with the 10.4 m Gran Telescopio Canarias (GTC) at the Observatorio del Roque de los Muchachos (ORM, La Palma, Spain). The spectra were obtained with OSIRIS (Optical System for Imaging and low-Intermediate-Resolution Integrated Spectroscopy) spectrograph in 9 one hour long observing blocks taken in four different nights: August 6, 2018, May 10 and 31 and June 2, 2019. The camera of the spectrograph consists of two Marconi CCD42-82 CCDs with a size of 2048 $\times$ 4096 pixels each and a 74 pixel gap between them. The physical size of the pixels is 15$\mu$m. We used a 2 $\times$ 2 binning, which provides a plate scale of 0.254 arcsec. OSIRIS was used in long-slit mode, using a slit 7.4 arcmin long and 0.8 arcsec wide. The slit center is placed in CCD2. We used a single spectrograph configuration for the observations. With the R1000B grism we cover from 3600 to 7750 \AA, practically the whole optical range and attaining an effective spectral resolution of 6.52 \AA. The total exposure time per object was 2646 s, divided in three consecutive 882 s-long exposures. This procedure permitted to accommodate the complete set of observations for each object in one-hour observing block. An additional 50 s exposure was taken for each object. This was done in order to avoid possible saturation of the brightest lines. In figures~\ref{fig:maps1} and \ref{fig:maps2} we show the apertures extracted for each object. The location and size of the apertures were selected to cover the brightest part of the nebula, trying to optimize the signal-to-noise ratio of the [\ion{N}{ii}] 5755 \AA\ line, which is necessary to determine the electron temperature (see Section~\ref{sec:ne_te}). 

In Table~\ref{tab:journal}, we present the coordinates of the centre of the aperture, Galactocentric distance of the nebula (see Section~\ref{sec:distances}), position angle (PA) of the slit, size of the aperture, mean airmass during the observations and mean observed -- uncorrected for reddening -- H$\alpha$ surface brightness in the aperture. We do not consider the effect of the atmospheric differential refraction because our targets are very extended objects, with angular diameters from several tens of arcseconds to arcminutes. The conditions of the ionized gas are not expected to change significantly in spatial scales of few arcseconds. 

The raw data were reduced using {\sc iraf}\footnote{{\sc iraf}, the Image Reduction and Analysis 
Facility, is distributed by the National Optical Astronomy Observatory, 
which is operated by the Association of Universities for Research 
in Astronomy under cooperative agreement with the National Science 
Foundation.} v2.16. We followed the standard procedures for the reduction of long-slit spectra. The wavelength calibration was performed making use of Hg-Ar, Ne and Xe lamps. We used observations of the standard stars GD140, GD153, Feige 110 and Ross 640 for flux-calibrating the spectra. The sky background emission shows some variation along the OSIRIS long-slit spectra and its removal requires doing it with care \citep{Fang:2015}. We extracted strips from the spectrograms free of emission of the nebula and multiplied them by a factor around one that makes the subtraction of the sky background satisfactory. In all objects except Sh 2-93, the emission of the nebula fills completely the CCD where the slit center is located (CCD2). In these cases, the sky emission strips were extracted from CCD1. 

The logarithmic reddening coefficient, $c$(H$\beta$), was calculated comparing the observed flux ratios of H$\alpha$, H$\gamma$ and H$\delta$ with respect to H$\beta$ and the theoretical ones determined by \cite{Storey:1995}. We have assumed the physical conditions given in  Section~\ref{sec:calculations} for each nebulae and followed an iterative process. We have assumed the reddening function, $f(\lambda)$, normalized to H$\beta$ derived by \cite{Cardelli:1989} considering $R_V$ = 3.1. In tables~\ref{TableA1} and \ref{TableA2} of Appendix~\ref{appendixA}, we present the dereddened line intensity ratios with respect to H$\beta$ of the emission lines identified in the spectrum of each nebulae.

\subsection{Distances using \textit{Gaia} DR2 parallaxes}
\label{sec:distances}

The $R_G$ values of our sample objects were calculated following the same methodology outlined in \citet{Mendez-Delgado:2020}, which is based on adopting the heliocentric distances derived by \citet{Bailer-Jones:2018} from \textit{Gaia} DR2 parallaxes \citep{gaiadr2} for the ionizing stars or the stellar content of the nebulae located at heliocentric distances smaller than 5 kpc, which is the case of all the objects observed in this work. As \citet{Mendez-Delgado:2020}, we adopt 8.2 $\pm$ 0.01 kpc for the distance of the Sun to the Galactic Centre. The details for each object are described below:

\textit{Sh\,2-29}. We estimate a heliocentric distance of $1.07^{+0.07}_{-0.06}$ kpc based on the ionizing star HD 165921 \citep{avedisovakondratenko84} (Gaia DR2 source 4066278846098710016). which corresponds to a Galactocentric distance of $R_G = 7.14^{+0.17}_{-0.16}$ kpc.

\textit{Sh\,2-32}. The estimation of the distance of this object is based on the B1V star CD$-$23 13997 (Gaia DR2 source 4066292830512043520), one of the two ionizing stars of the region \citep{avedisovakondratenko84}. The resulting heliocentric distance is 1.59$^{+0.15}_{-0.13}$ kpc, while $R_G = 6.63^{+0.25}_{-0.22}$ kpc.

\textit{Sh\,2-47}. We use the B0.5III ionizing star BD$-$15~4913 \citep{avedisovakondratenko84} (Gaia DR2 source 4098093077571330816). Its  heliocentric distance is $1.66 ^{+0.14}_{-0.12}$ kpc, which corresponds to $R_G = 6.62^{+0.24}_{-0.27}$ kpc.

\textit{Sh 2-48}. Our distance estimation is based on the parallax of the ionizing O8V star BD$-$14 5014 \citep{avedisovakondratenko84} (Gaia DR2 source 4098272263591120128). This source is located at a heliocentric distance of $4.35^{+2.09}_{-1.22}$ kpc, which corresponds to $R_G = 4.22^{+2.19}_{-1.32}$ kpc.

\textit{Sh 2-53}. This region is related to the complex of \ion{H}{II} regions associated to the SNR G18.1-0.1 system. \citet{Paron:2013} indicated three O-type stars as candidates for the ionizing sources of the \ion{H}{II} regions. These stars are presented in their Table 2 as S3, S5 and S15, which correspond to the Gaia DR2 sources 4152517112126036992, 4152566800618412928 and 4152566628819783424, respectively. The heliocentric distances of these sources given by the $Gaia$ DR2 parallaxes are $8.11^{+2.71}_{-1.88}$ kpc, $4.36^{+1.70}_{-1.08}$ kpc and $3.36^{+1.64}_{-0.91}$ kpc, respectively. Previous kinematic studies reported heliocentric distances between 3.8 kpc and 4.5 kpc \citep{Paladini:2003, Kassim:1989, Yadigaroglu:1997}. We discard the Gaia DR2 source 4152517112126036992 (S3 from \citealt{Paron:2013}) since it might not be associated with the region due to its very discrepant parallax. We adopt an average distance based on the remaining two stars, obtaining $R_G = 4.7\pm1.1$ kpc.

\textit{Sh 2-54}. \citet{GonzalezRosado84} place the Wolf-Rayet star WR\,113 in Sh 2-54, which corresponds to the Gaia DR2 source 4153716198275554176, located 1.82$^{+0.19}_{-0.16}$ kpc away from the Sun. This value is in agreement with previous kinematic determinations by \citet{Blitz:1982} and \citet{Forbes:2000}, who obtain heliocentric distances of $1.9\pm 0.3$ kpc and $2.0 \pm 0.2$ kpc, respectively. The Gaia parallax analysed by \citep{Bailer-Jones:2018} implies a $R_G = 6.51^{+0.28}_{-0.25}$ kpc.

\textit{Sh 2-88B}. \citet{Lortet1974} identified two bright H$\alpha$-knots associated to Sh\,2-88, designating them as Sh\,2-88A and Sh\,2-88B (see Fig.~\ref{fig:maps2}). \citet{avedisovakondratenko84} reported three ionizing stars for Sh\,2-88: HD 338916, HD 338926 and LS II $+$25 09. Such sources are identified in the Gaia DR2 catalog as 2020947043840246144, 2020923679216984064 and 2020883959358967552, with heliocentric distances of $1.93^{+0.16}_{-0.14}$ kpc,  $2.08^{+0.19}_{-0.16}$ kpc and $4.38^{+2.98}_{-1.89}$ kpc, respectively \citep{Bailer-Jones:2018}. We use the weighted average value of these objects, $2.00\pm0.12$ kpc, to derive $R_G = 7.45\pm0.18$ kpc. 

\textit{Sh 2-93}. There is a lack of information about the ionizing sources of this \ion{H}{II} region. \citet{Forbes89} identified an O9.5V exciting star, Anon 1, at the center of the region (see his Figure 1). However, we were not able to identify this star in the Gaia DR2 catalog. Instead, we seek for the distances of all stars within a radius of 5 arcmins centred on Sh~2-93. This search should cover the stellar content of the star cluster [BDS2003]~16, associated with Sh~2-93 \citep{bica03}. The resulting distribution of distances give us a median value of 2.53 kpc with a deviation represented by the 84th and 16th percentile of 0.45 kpc and 0.55 kpc, respectively, which implies $R_G = 7.45^{+0.41}_{-0.47}$ kpc. The adopted heliocentric distance is in agreement with previous estimates between 2.6 and 3.7 kpc \citep{huntermassey90,Paladini:2003}.

\begin{table*} 
\centering \caption{Data of the sample objects and their observations.} 
\label{tab:journal} 
\begin{tabular}{lccccccc} 
\hline 
& R.A.$^{\rm a}$ & Decl.$^{\rm a}$ & $R_\mathrm{G}^{\rm b}$ & PA & Extracted area & &  \\ 
{\ion{H}{ii}} Region & (J2000) & (J2000) &  (kpc) &  ($^\circ$) & ($\mathrm{arcsec^2}$) &Airmass & $S$(H$\alpha$)$^{\rm c}$\\ 
\hline 
Sh 2-29& 18:09:49.90 & $-$24:04:14.5 & 7.14 $\pm$ 0.17 & 149 & 20.3 $\times$ 0.8 & 1.66 & 1.22 \\ 
Sh 2-32 & 18:09:52.72 & $-$23:38:39.5 & 6.63 $\pm$ 0.24 & 90 & 26.7 $\times$ 0.8 & 1.65 & 0.68 \\ 
Sh 2-47 & 18:18:16.02 & $-$15:36:17.2 & 6.62 $\pm$ 0.23 & 75 & 25.4 $\times$ 0.8 & 1.41 & 0.28 \\ 
Sh 2-48 & 18:22:12.46 & $-$14:36:19.1 & 4.2$^{+2.2}_{-1.3}$ & 103 & 25.4 $\times$ 0.8 & 1.45 & 0.36 \\ 
Sh 2-53 & 18:25:16.21 & $-$13:09:12.2 & 4.7 $\pm$ 1.1 & 46 & 35.6 $\times$ 0.8 & 1.49 & 0.28 \\ 
Sh 2-54 & 18:17:48.99 & $-$11:43:45.4 & 6.51 $\pm$ 0.27 & 90 & 22.9 $\times$ 0.8 & 1.41 & 2.36 \\ 
Sh 2-82 & 19:30:22.80 & 18:15:14.4 & 7.73 $\pm$ 0.17 & 149 & 50.8 $\times$ 0.8 & 1.14 & 0.20 \\ 
Sh 2-88B& 19:46:46.51 & 25:12:39.4 & 7.45 $\pm$ 0.18 & 62 & 26.7 $\times$ 0.8 & 1.21 & 1.86\\ 
Sh 2-93 & 19:54:58.17 & 27:12:45.3 & 7.45 $\pm$ 0.44 & 8 & 17.8 $\times$ 0.8 & 1.39 & 1.33 \\ 
\hline 
\end{tabular} 
\begin{description} 
\item[$^{\rm a}$] Coordinates of the centre of the extracted aperture. 
\item[$^{\rm b}$] Galactocentric distances assuming the Sun at 8.2 kpc. 
\item[$^{\rm c}$] Measured mean H$\alpha$ surface brightness in the areas extracted for spectroscopical analysis in units of 10$^{-14}$ erg cm$^{-2}$ s$^{-1}$ arcsec$^{-2}$.
\end{description} 
\end{table*} 

\subsection{Additional sample of Galactic \ion{H}{II} regions}
\label{sec:add_sample}

We have included the sample of Galactic \ion{H}{ii} regions compiled by \cite{Arellano-Cordova:2020a}, which covers a range of $R_G$ from 6 to 17 kpc determined with \textit{Gaia} DR2 parallaxes for most of the objects. The data  comprise deep high or intermediate spectral resolution spectra  \citep{Esteban:2004, Esteban:2014, Esteban:2018, Garcia-Rojas:2014, Fernandez-Martin:2017}, that permit to obtain direct measurement of electron temperature, $T_{\rm e}$, and an accurate computation of chemical abundances of the nebulae. 
 
\citet{Arellano-Cordova:2020a} performed an improved and homogeneous analysis of the chemical abundances using updated atomic data and some considerations on the temperature structure of the nebula. For this paper, we have used the N and O abundances obtained by \cite{Arellano-Cordova:2020a}.

\begin{figure} 
\begin{center}
    \includegraphics[width=0.4\textwidth, trim=30 0 30 0,  clip=yes]{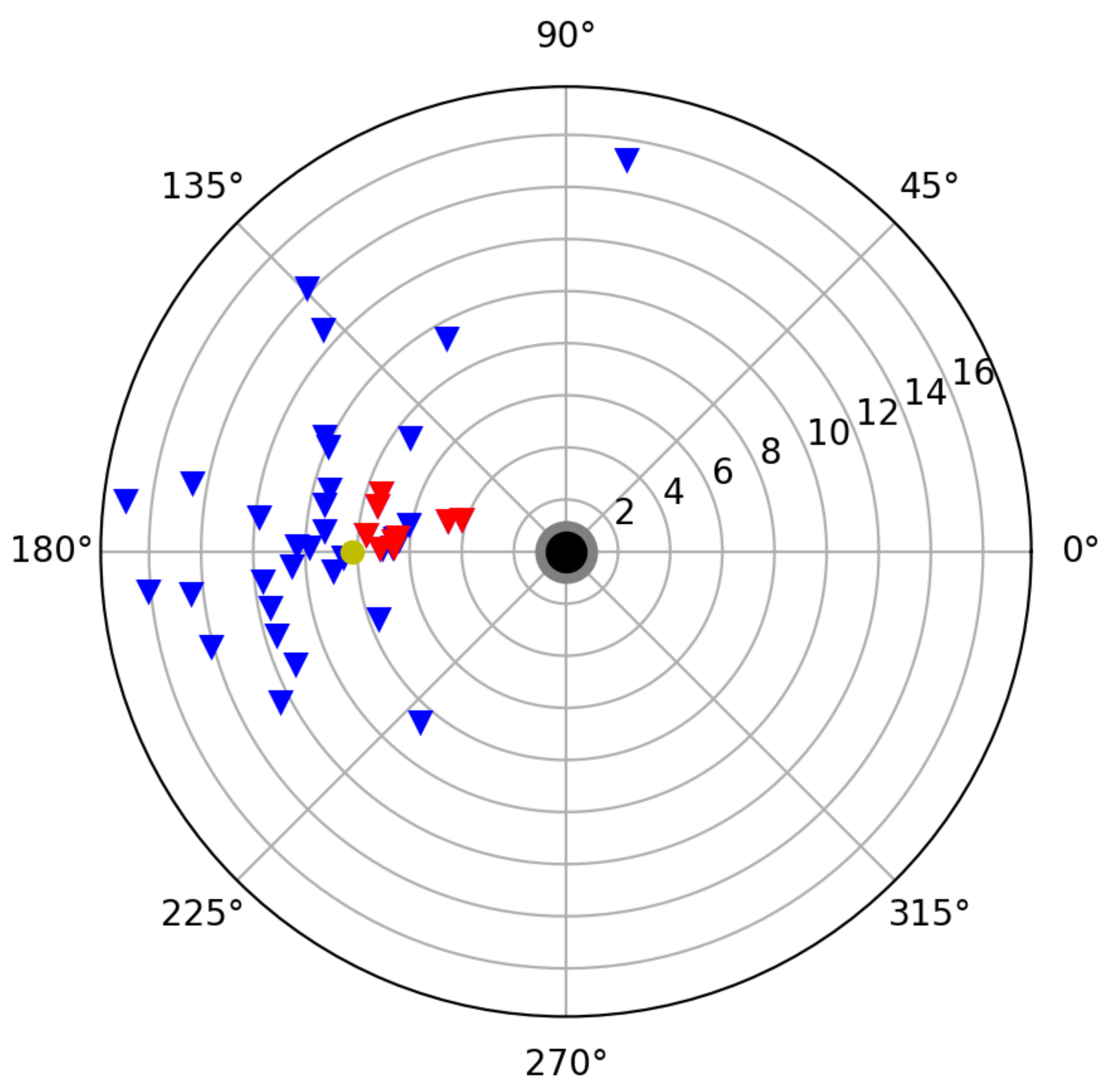}
    \caption{Spatial distribution of 42 \ion{H II} regions on to the Galactic plane with respect to the centre of the Milky Way. Red symbols show the nine \ion{H II} regions presented this work and blue ones those of the additional sample taken from \citet{Arellano-Cordova:2020a}. The yellow circle represents the position of the Sun at 8.2 kpc from the Galactic centre. The concentric circles indicate increasing Galactocentric distances (in kpc).}
    \label{fig:sample}
   \end{center}
    \end{figure}

 The spatial distribution of our sample of 42 \ion{H}{ii} regions is represented in Fig~\ref{fig:sample}. The red symbols indicate the 9 objects observed in this work and the blue symbols represent the additional sample.

\section{physical conditions and chemical abundances}
\label{sec:calculations}

We have used the Python package \textsc{PyNeb} version 1.1.5  \citep{Luridiana:2015} and the atomic data set indicated in  Table~\ref{tab:atomic_data} to determine the physical conditions and chemical abundances of the nebulae.

 \begin{table*}\footnotesize
 \caption{Atomic data used in this work for CELs}
 \begin{center}
 \begin{tabular}{lcc}
 \hline
 \multicolumn{1}{l}{Ion} & \multicolumn{1}{c}{Transition Probabilities} &
 \multicolumn{1}{c}{Collision Strengths} \\
 \hline

O$^{+}$   &  \citet{FFT:2004} & \citet{Kisielius:2009}\\
O$^{2+}$  &  \citet{Wiese:1996}, \citet*{Storey:2000} & \citet{Storey:2014}\\
N$^{+}$   &  \citet{FFT:2004} & \citet{Tayal:2011}\\
S$^{+}$   &  \citet{Podobedova:2009} & \citet{Tayal:2010}\\
Cl$^{2+}$ &  \citet{Fritzsche:1999} & \citet{Butler:1989}\\
Ar$^{2+}$ &   \citet{Mendoza:1983}, \citet{Kaufman:1986}  & \citet*{Galavis:1995}\\
Fe$^{2+}$ & \citet{Quinet96}, \citet{Johansson00} & \citet{Zhang96}\\
 \hline
 \end{tabular}
 \end{center}
 \label{tab:atomic_data}
 \end{table*}

\subsection{Density and electron temperature}
 \label{sec:ne_te}

 The electron density, $n_{\rm{e}}$, was calculated using the  [\ion{S}{ii}] $\lambda6717/\lambda6731$ and/or  [\ion{Cl}{iii}] $\lambda5518/\lambda5538$ line intensity ratios. For \ion{H}{ii} regions in the low-density regime, i.~e. $n_{\rm{e}} <100$ cm$^{-3}$, we assumed $n_{\rm{e}}=100$ cm$^{-3}$ \citep{Osterbrock:2006} for the calculations. Although we could measure or put an upper limit to $n_{\rm{e}}$([\ion{Cl}{iii}]) in Sh\,2-29, Sh\,2-54 and Sh\,2-88B, we discard the use of this diagnostic in our calculations owing to the large uncertainties involved. 
 
 For the whole sample, we calculate $T_{\rm e}$([\ion{N}{ii}]) using the [\ion{N}{ii}]~($\lambda6548+\lambda6584)/\lambda5755$ line intensity ratio. Although the [\ion{N}{ii}] $\lambda5755$ line intensity may be affected by some contribution from N$^{2+}$ recombination  (\citealt{Rubin:1986,Stasinska:2005, Liu:2000}), this effect has been demonstrated to be very small or even negligible in \ion{H}{ii} regions \citep[see e.~g.][]{Garcia-Rojas:2007a, Esteban:2018}. Moreover, the fact that the correction is rather uncertain and that our nebulae have very low ionization degree with most of the N is in the form of N$^+$, have led us to not correct from this contribution. 
 We have to emphasize that although some of these objects (Sh\,2-48, Sh\,2-54, Sh\,2-88B and Sh\,2-93) have been spectroscopically observed in previous works \citep[see e.~g.][]{Blitz:1982, Deharveng:2000, Caplan:2000}, this is the first time that optical diagnostics have been used to compute the electron temperature and chemical abundances, avoiding the biases discussed in Section~\ref{sec:intro}.
We have also computed $T_{\rm e}$([\ion{S}{ii}]) for Sh\,2-29, Sh\,2-32, Sh\,2-54, and Sh\,2-88B  and  $T_{\rm e}$([\ion{O}{ii}]) for Sh\,2-54, Sh\,2-88B, and Sh\,2-93 using the [\ion{S}{ii}]~($\lambda4069+\lambda4076)/(\lambda6717 +\lambda6731$) and [\ion{O}{ii}]~($\lambda3726+\lambda3729)/(\lambda\lambda7319,20 + \lambda\lambda7330,31)$ line intensity ratios, respectively. We have ignored the recombination contribution to the trans-auroral [\ion{O}{ii}] $\lambda\lambda7319,20 + \lambda\lambda7330,31$ lines for the same reasons described above for [\ion{N}{ii}] $\lambda5755$. $T_{\rm e}$([\ion{O}{iii}]) has only been computed from the [\ion{O}{iii}]~($\lambda4959+\lambda5007)/\lambda4363$ line intensity ratio for Sh\,2-54. 
%

 The effect of using the $T_{\rm e}$([\ion{N}{ii}]) - $T_{\rm e}$([\ion{O}{iii}]) temperature relation in chemical abundance determinations was assessed by \citet{Arellano-Cordova:2020b} for a sample of extragalactic \ion{H}{ii} regions. These authors evaluated the differences between using a temperature relation and both $T_{\rm e}$([\ion{N}{ii}]) and $T_{\rm e}$([\ion{O}{iii}]) to compute the chemical abundances of N and O. \citet{Arellano-Cordova:2020b} showed that the temperature relation by \citet{Campbell:1986}, whose relation is similar to that obtained by \citet{Esteban:2009} using observational data, provided the best results, but the differences in the abundances of N and O can reach up to 0.2 dex. These authors also concluded that the use of a single electron temperature, $T_{\rm e}$([\ion{N}{ii}]) or $T_{\rm e}$([\ion{O}{iii}]), can lead to even larger differences than when using both temperatures. However, \citet{Arellano-Cordova:2020b} discarded using Galactic \ion{H}{ii} regions in their analysis since these observations cover only a small part of the nebula. In our study, we use the temperature relation proposed by \citet{Esteban:2009} to estimate $T_{\rm e}$([\ion{O}{iii}]) from $T_{\rm e}$([\ion{N}{ii}]) only for objects with very low ionization degree (O$^{2+}$/O$<$0.2) where the auroral \ion{O}{iii} $\lambda$4363 line has not been detected and that comprises 50\% of the sample. There are three exceptions to this rule: Sh\,2-90 and Sh\,2-48 where O$^{2+}$/O ratios are slightly higher (0.3 and 0.35 respectively) and NGC\,2579, with O$^{2+}$/O=0.62, where the undetected line was the auroral [\ion{N}{ii}] $\lambda$5755 line and the relation was used to compute $T_{\rm e}$([\ion{N}{ii}]) \citep{Esteban:2013}. In general, this approximation is expected not to bias the results of O/H and N/H ratios as for most of these objects, O$^+$ and N$^+$ strongly dominate over other ionization stages.
The final results of the physical conditions computed for each nebulae are included in Table~\ref{tab:phy-abundances}. 

We have only considered $T_{\rm e}$([\ion{N}{ii}]) to characterize electron temperature in the low ionization zones of the nebulae. There are two reasons for this. Firstly, the uncertainties estimated for $T_{\rm e}$([\ion{N}{ii}]) are significantly lower than those derived for $T_{\rm e}$([\ion{O}{ii}]) and $T_{\rm e}$([\ion{S}{ii}]). Secondly, $T_{\rm e}$([\ion{O}{ii}]) and $T_{\rm e}$([\ion{S}{ii}])  diagnostics may be affected by some observational and physical problems. $T_{\rm e}$([\ion{O}{ii}]) is highly dependent on $n_{\rm{e}}$, and the [\ion{O}{ii}]~$(\lambda\lambda7319+20,7330+31)$ lines can be contaminated by telluric emission. On the other hand, the suitability of including $T_{\rm e}$([\ion{S}{II}]) as representative of the temperature of the low ionization zone is questionable. S$^+$ has an ionization potential (IP) lower than those of H$^+$, O$^+$ or N$^+$ and part of the zone were S$^+$ lies may be outside the ionization front, where the physical conditions of the gas may be different. Finally, we have adopted $T_{\rm e}$([\ion{O}{iii}]) for the high ionization zone.

 \begin{table*}
 \caption{Physical conditions and abundances}
 \begin{tabular}{l l l l l l l l l l l}
 \hline
 \multicolumn{1}{l}{} & \multicolumn{1}{l}{Sh 2-29} &  \multicolumn{1}{l}{Sh 2-32}  &  \multicolumn{1}{l}{Sh 2-47}  &  \multicolumn{1}{l}{Sh 2-48}
 & \multicolumn{1}{l}{Sh 2-53} & \multicolumn{1}{l}{Sh 2-54} & \multicolumn{1}{l}{Sh 2-82} & \multicolumn{1}{l}{Sh 2-88B} & \multicolumn{1}{l}{Sh 2-93}\\
 \hline

\multicolumn{10}{c}{Physical conditions$^{a}$} \\
$n_{\rm e}$([\ion{S}{ii}])     &  100$\pm$40                  & 100$\pm$50                  & 100$\pm$70                            &  100$\pm$60                        & 300$\pm$100                & 200$\pm$50               & 100$\pm$70                     & 900$\pm$100                   & 500$\pm$50              \\
$n_{\rm e}$([\ion{Cl}{iii}])   &  700$^{+3000}_{-600}$        & $-$                         & $-$                                   &  $-$                               & $-$                        & 200$^{+900}_{-100}$      & $-$                            &  < 100                        & $-$                     \\
\\                                                                                                                                                                                                                               
                                                                                                                                                                                                                                 
$T_{\rm e}$([\ion{N}{ii}])    & 7500$\pm$200                & 7100$\pm$200                & 7000$\pm$500                            & 6800$\pm$500                       & 7000$\pm$400               & 7500$\pm$200              & 7000$\pm$900                   & 8300$\pm$200                 & 7900$\pm$300            \\
$T_{\rm e}$([\ion{O}{ii}])    & $-$                         & $-$                         & $-$                                     & $-$                                & $-$                        & 7200$^{+500}_{-400}$      & $-$                            & 8500$\pm$400                 & 8900$^{+700}_{-600}$       \\                                                                                                                   
$T_{\rm e}$([\ion{S}{ii}])    & 8600$\pm$500                & 8300$\pm$900                & $-$                                     & $-$                                & $-$                        & 7300$^{+1800}_{-1100}$    & 14700$^{+6800}_{-4300}$        & 21100$^{+8500}_{-5200}$      & $-$          \\
$T_{\rm e}$([\ion{O}{iii}])   & 6300$\pm$200$^{b}$      & 5800$\pm$300$^{b}$      & 5600$^{+600}_{-700}$$^{b}$          & 5300$\pm700$$^{b}$             & 5600$\pm$600$^{b}$     & 8800$^{+800}_{-1000}$     & 5600$\pm$1200$^{b}$        & 7300$\pm$300$^{b}$       & 7000$\pm400$$^{b}$      \\
\\

\multicolumn{10}{c}{Ionic abundances$^{c}$} \\
He$^+$                         & 10.52$\pm$0.01   & $-$ & 9.92$\pm$0.03    &  10.87$\pm$0.07  &  10.59$\pm$0.04  & 10.95$\pm$0.03  & 10.14$\pm$0.02  &  10.67$\pm$0.04   & 10.47$\pm$0.02            \\

C$^{2+}$                      & $-$                              &  $-$                            &  $-$                            & $-$                           & $-$                            &  8.49$\pm$0.11                    &  $-$                     &                               & $-$                  \\
N$^+$                         & 7.85$\pm$0.05                    &  7.79$\pm$0.07                  &  7.84$^{+0.19}_{-0.13}$        &  7.75$^{+0.19}_{-0.15}$        &  8.06$\pm$0.15                 &  7.56$\pm$0.06                    &  7.81$^{+0.30}_{-0.19}$  &  7.66$\pm$0.05                &  7.86$\pm$0.08   \\        
O$^+$                         & 8.62$\pm$0.07                    &  8.53$\pm$0.11                  &  8.68$^{+0.31}_{-0.22}$         & 8.51$^{+0.34}_{-0.24}$        & 8.73$\pm$0.26                  & 8.34$\pm$0.11                     & 8.60$^{+0.49}_{-0.31}$          & 8.47$\pm$0.09          & 8.41$\pm$0.14   \\                      
O$^{2+}$                      & 7.46$\pm$0.12                    &  6.90$\pm$0.17                  &  7.90$^{+0.55}_{-0.35}$  &    8.24$^{+0.52}_{-0.42}$            &    7.85$^{+0.44}_{-0.34}$  &    7.65$^{+0.33}_{-0.19}$  &    $-$                     &    7.67$\pm$0.12           &    7.25$\pm$0.20           \\
S$^+$                         & $  6.76 ^{+ 0.05 }_{- 0.04 }$    &  $ 6.74 ^{+ 0.06 }_{- 0.07 }$   &  $ 6.76 ^{+ 0.18 }_{- 0.14 }$  &  $ 6.48 ^{+ 0.16 }_{- 0.14 }$  &  $ 6.56 ^{+ 0.19 }_{- 0.14 }$  &  $ 6.16 ^{+ 0.08 }_{- 0.06 }$     &  $ 6.72 ^{+ 0.30 }_{- 0.19 }$   &    $  6.23 \pm 0.05 $             &    $ 6.53\pm0.08$ \\
S$^{2+}$                      & $  6.69\pm0.20$                  &  $-$                            &  $ -$                           &  $ -$                         &  $ -$                          &  $ 6.71 ^{+ 0.17 }_{- 0.28 }$     &  $ -$                           &    $  6.92^{+ 0.11 }_{- 0.15 }$    &    $ 6.76^{+ 0.16 }_{- 0.28 }$ \\

Cl$^{2+}$                     & $  4.69 \pm 0.12 $               &  $-$                           &  $-$                           &  $-$                           & $-$                            &  $ 5.05 \pm0.09$                  &  $-$                           &    $ 4.87 ^{+ 0.10 }_{- 0.13 }$    &      $-$                         \\    

Ar$^{2+}$                     & $ 5.80 ^{+ 0.05 }_{- 0.09 }$     &  $ 5.03 ^{+ 0.11 }_{- 0.16 }$  & $ 5.06 ^{+ 0.26 }_{- 0.25 }$ &  $ 6.21 ^{+ 0.36 }_{- 0.14 }$    &  $ 5.92 ^{+ 0.19 }_{- 0.18 }$  &  $ 6.04 ^{+ 0.08 }_{- 0.14 }$    &  $ 5.47 ^{+ 0.31 }_{- 0.49 }$  &   $ 6.02 ^{+ 0.08 }_{- 0.05 }$  & $ 5.88 ^{+ 0.15 }_{- 0.09 }$ \\
Fe$^{2+}$ & 5.4: &  $-$ & $-$ &$-$ &$-$ & 5.58$\pm$0.05 & $-$ &  $6.05^{+0.52} _{-0.36}$& $-$\\

\\

\multicolumn{10}{c}{Total abundances$^{d}$} \\
C                             &  $-$                 &  $-$                &   $-$                 &   $-$        &  $-$        &  8.78$^{+0.35}_{-0.28}$        &  $-$                         &  $-$                 &  $-$  \\

N                             &   7.87$\pm$0.10                  &   7.80$\pm$0.11                 &   7.86$^{+0.30}_{-0.26}$             &   8.02$^{+0.39}_{-0.29}$        &  8.13$^{+0.30}_{-0.26}$        &  7.66$^{+0.14}_{-0.11}$         &  7.84$^{+0.26}_{-0.22}$$^{e}$      &  7.72$\pm$0.12                  &  7.88$\pm$0.15  \\

O                             &   8.65$\pm$0.07                  &   8.54$\pm$0.11                 &  8.65$^{+0.30}_{-0.24}$              &   8.69$^{+0.38}_{-0.28}$        &  8.78$^{+0.27}_{-0.24}$        &  8.43$\pm$0.11                  &  8.60$^{+0.49}_{-0.31}$      &  8.54$\pm$0.08                 &  8.44$\pm$0.13                        \\

S                             &   $7.02 ^{+ 0.08 }_{- 0.11 }$    &   $-$                            &  $ -$                                &   $ -$                          &  $ -$                          &   $ 6.81 ^{+ 0.20 }_{- 0.12 }$  &   $-$                       &  $ 7.0 ^{+ 0.11 }_{- 0.09 }$    & $ 6.96 ^{+ 0.15 }_{- 0.18 }$ \\

Cl                            &   $-$                  &   $-$                           &  $-$                                 &   $-$                           &  $-$                           &  $ 5.15\pm0.07$                 &   $-$                        &  $ -$  & $-$  \\

Ar                            &   $-$               &   $ - $             &  $ -$                   &   $ 6.32^{+0.31}_{-0.18}$             &  $- $              &   $-$&   $-$                        &  $ -$             & $ - $ \\


\\

\hline                     
\end{tabular}
\label{tab:phy-abundances}
\begin{description} 
\item[]Note: The total abundances of S, Cl and Ar are not reported for the objects outside the validity range of the ICF schemes used \citep[see][]{Arellano-Cordova:2020a}. Colons indicate an uncertainty of the order or greater than 40 per cent.
\item[]$^{a}$$n_{\rm e}$ in units cm$^{-3}$; $T_{\rm e}$ in K
\item[]$^{b}$ Estimated from $T_{\rm e}$(\ion{N}{ii}) using equation (3) of \citet{Esteban:2009}
\item[]$^{c}$ in units of 12+log(X$^{\rm n+}$/H$^{+}$) 
\item[]$^{d}$ in units of 12+log(X/H) 
\item $^{e}$ N abundances determined without ICF scheme. 

\end{description}
\end{table*}
 \label{tab:phy-abundances}

\subsection{Chemical abundances}
\subsubsection{Ionic abundances}
\label{sec:ionic}

We adopted $T_{\rm e}$([\ion{N}{ii}]) to calculate the abundances of O$^{+}$, N$^{+}$, S$^{+}$, Fe$^{2+}$ and Cl$^{2+}$ and $T_{\rm e}$([\ion{O}{iii}]) for He$^{+}$, O$^{2+}$, C$^{2+}$ and Ar$^{2+}$.
We used the [\ion{O}{ii}] $\lambda\lambda3726, 3729$, [\ion{N}{ii}] $\lambda\lambda6548, 6584$ and [\ion{S}{ii}] $\lambda\lambda6717, 6731$  line intensities to calculate the O$^{+}$, N$^{+}$ and S$^{+}$ abundances, respectively. In the case of the O$^{2+}$ abundances, we used the line intensities of [\ion{O}{iii}] $\lambda\lambda4959,5007$ for all the objects of our sample except Sh\,2-82 because [\ion{O}{iii}] lines were not detected in this nebula. We calculate the ionic abundances of Cl$^{2+}$ using the intensities of [\ion{Cl}{iii}] $\lambda\lambda5518, 5538$ lines for Sh\,2-29, Sh\,2-54 and Sh\,2-88B. The Ar$^{2+}$ ionic abundance was computed using the [\ion{Ar}{iii}] $\lambda\lambda7135$ line for six of the objects, while for Sh\,2-54, Sh\,2-88B and Sh\,2-93 we used the [\ion{Ar}{iii}] $\lambda7135,7751$ lines. The Fe$^{2+}$ abundances were calculated using either the [\ion{Fe}{III}] $\lambda4658$ or [\ion{Fe}{III}] $\lambda5271$ lines for Sh\,2-29, Sh\,2-54 and Sh\,2-88B. 

We determine the He$^{+}$ abundances for eight objects in our sample by using the available intensities from the following recombination \ion{He}{I} lines: $\lambda$4026, $\lambda$4388, $\lambda$4471, $\lambda$4922, $\lambda$5876 and $\lambda$6678 \citep{Mendez-Delgado:2020} and the effective recombination coefficients of \citet{Porter12,Porter13}. The resulting values, shown in Table \ref{tab:phy-abundances}, are the average abundances obtained with each available line, weighted by the inverse of the square of their  uncertainties. Finally, we have detected the C\thinspace II $\lambda$4267 RL in the spectrum of Sh~2-54. We determined the C$^{2+}$ abundance assuming case B, $T_{\rm e}$([\ion{O}{iii}]) and using the  
effective recombination coefficients calculated by \citet{daveyetal00}.

\subsubsection{Total abundances}

In principle, for the normal conditions of \ion{H}{ii} regions, the total O abundance is determined by adding the contribution of the O$^{+}$ and O$^{2+}$ abundances. For the rest of the elements, we need to use ICFs to take into account the contribution of unseen ions. \citet{Arellano-Cordova:2020a} analyzed the behavior of some of the most used ICF schemes in the literature for determining the abundance of different elements in Galactic \ion{H}{ii} regions. Among the sets of ICFs considered by \citet{Arellano-Cordova:2020a} there are the new schemes from Amayo et al. (in preparation), which show the best general behavior. The ICFs of 
 Amayo et al. are based on a detailed selection of photoionization models computed  by \citet{Vale-Asari:2016}. Those ICFs depend on the ionization degree of the nebulae, parameterized by O$^{2+}$/O and are valid for the whole range of typical values of that parameter in \ion{H}{ii} regions. Besides, Amayo et al.  provide analytical expressions to compute the uncertainties associated with each ICF. 

In particular, we use the ICF of Amayo et al. to calculate the total abundance of C and N. For the rest of the elements, we adopt the ICFs adopted by  \citet{Arellano-Cordova:2020a} in the sake of consistency with the results of that paper. Most of the objects studied in this paper show and ionization degree outside the validity range of the ICF schemes used for determining the total S, Cl and Ar abundances. Therefore, the abundances of those elements have not been calculated. We do not estimate the total abundance of He since the contribution of the He$^{0}$/H$^{+}$ depends essentially on the effective temperature of the ionizing source \citep{Delgado-Inglada:2014} which requires further study and is beyond the purposes of this work. The wide range of values of He$^{+}$/H$^{+}$ does not mean a wide range of total He abundances but rather different contributions of neutral helium within the ionized gas.

In general, the ionization degree of the sample is rather low and so is the expected contribution of the N$^{2+}$/H$^{+}$ ratio to the total N abundance. In the case of Sh\,2-82, its values of N/H and N/O are computed without the use of an ICF due to its very low ionization degree, reflected in the absence of [\ion{O}{iii}] $\lambda\lambda$4959,5007 lines, similarly to some of the objects observed by \citet{Esteban:2018}.

Sh\,2-54 is the only object for which we have determined the C/H ratio from the intensity of a \ion{C}{ii} RL. We would also need to have the O/H determined from RLs to compute a suitable C/O ratio for this object. However, we do not detect RLs of any ion of O in its spectrum. 
Following the procedure of \citet{Arellano-Cordova:2020a}, we can make a rough estimation of the C/O ratio of Sh\,2-54 applying an average empirical correction of O/H$_{\rm (RLs)}$ $-$ O/H$_{\rm (CELs)} = 0.19\pm0.05$ dex to the O/H ratio determined form CELs for this object. This factor corrects for the average effect of the abundance discrepancy \citep{Garcia-Rojas:2007a} in Galactic \ion{H}{ii} regions. The C/O ratio we obtain for Sh\,2-54 is the highest value found for Galactic \ion{H}{II} regions, log(C/O) = 0.16 $\pm$ 0.40 dex, but also very uncertain \citep[see][]{Arellano-Cordova:2020a}.

For the additional sample of Galactic \ion{H}{ii} regions, we have taken the values of O/H, N/H and N/O abundance ratios obtained by \citet{Arellano-Cordova:2020a}. In 12 of those objects, their N abundance were calculated without using an ICF scheme due to their low ionization degree. For the rest of the nebulae, we have considered the N abundances obtained by \citet{Arellano-Cordova:2020a} using the ICF of Amayo et al.

The uncertainties associated with $n_{\rm e}$, $T_{\rm e}$ and the ionic and total abundances were calculated following the procedure followed by \citet{Arellano-Cordova:2020a}, which is based on Monte Carlo simulations. We generate 500 random values for each line intensity ratio assuming a Gaussian distribution with a sigma equal to the associated uncertainty. 
In Table~\ref{tab:phy-abundances}, we include the results for the ionic abundances of He$^{+}$, C$^{2+}$, N$^{+}$, O$^{+}$, O$^{2+}$, S$^{+}$, Cl$^{2+}$, Ar$^{2+}$ and Fe$^{2+}$ and the total abundance of C, N, O, S, Cl and Ar for our sample of nine Galactic \ion{H}{ii} regions. 
 

\section{The radial abundance gradients of O and N}
\label{sec:abundance-gradients}

We perform least-squares linear fits to compute the radial abundance gradient of O/H, N/H and N/O ratios using the nine Galactic \ion{H}{ii} regions studied in this paper and the additional sample of object.Both samples were analyzed using the same procedure and atomic data set.
The computation of the gradients includes the uncertainties in the abundances and $R_G$ using Monte Carlo simulations. We have generated 10$^4$ random values of the O and N abundances and $R_G$ assuming a Gaussian distribution with a sigma equal to the uncertainties estimated for each quantity. The uncertainties associated to the slope and intercept of the linear fits correspond to the standard deviations of the values of these two parameters obtained from the fits \citep{Esteban:2018}.

\subsection{The radial gradient of O/H}
\label{sec:oxygen}
\begin{figure*} 
\begin{center}
    \includegraphics[width=0.9\textwidth, trim=30 0 30 0,  clip=yes]{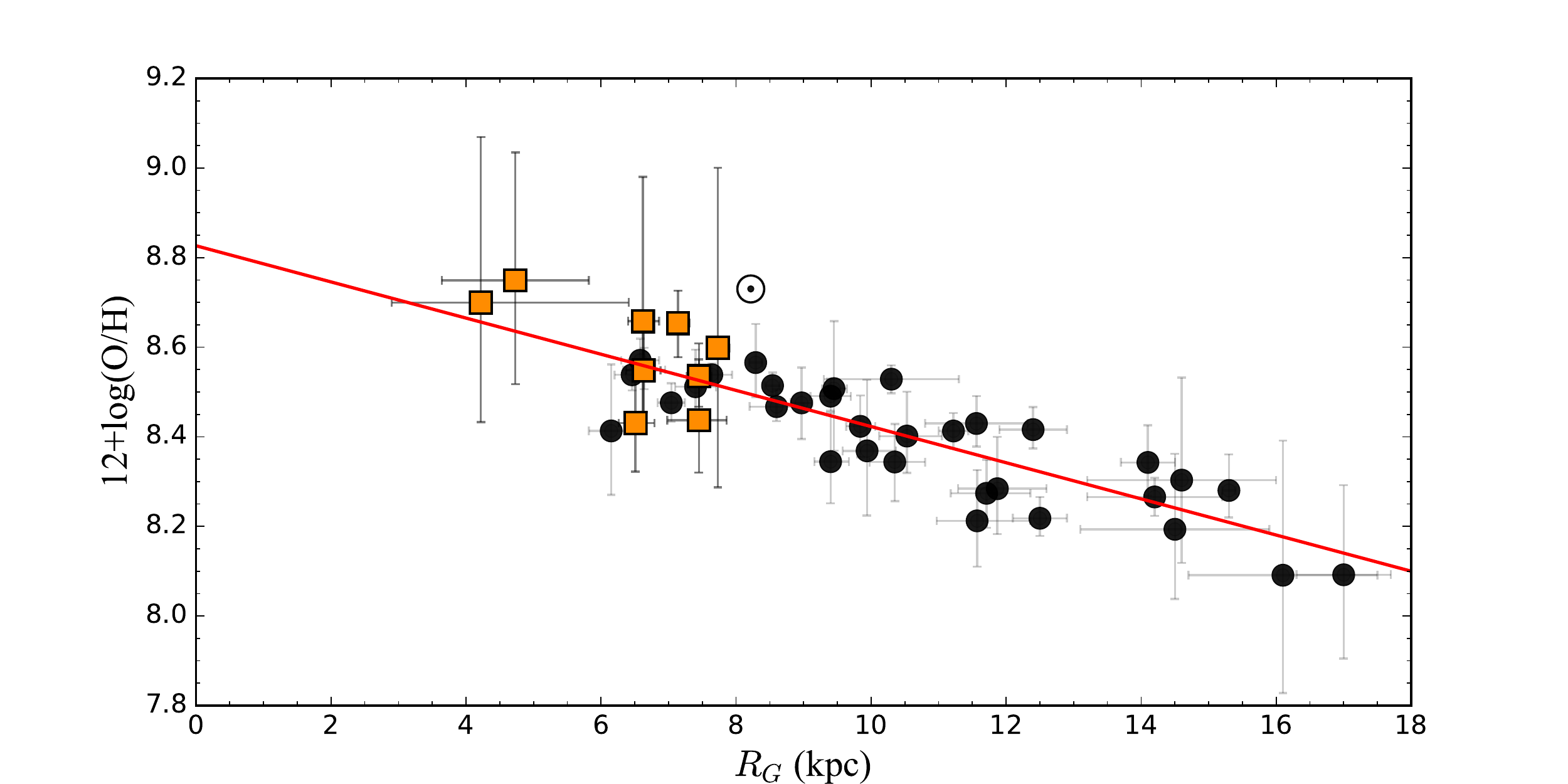}
    \caption{The radial abundance gradient of O for a sample of 42 Galactic \ion{H}{ii} regions as a function of the Galactocentric distance ($R_G$). The orange squares represent the new observations presented in this work and the circles the additional sample compiled by \citet{Arellano-Cordova:2020a}. The line indicates the least-squares fit to the data. The Solar symbol indicates the photosphere O abundance of the Sun given by \citet{Lodders:2019}.}
    \label{fig:O-gradient}
   \end{center}
    \end{figure*}

In Fig. \ref{fig:O-gradient}, we present the radial abundance gradient of oxygen using 42 Galactic \ion{H}{ii} regions with direct determinations of $T_{\rm e}$ and covering $R_G$ values from 4 to 17 kpc.  
 From the least-squares linear fit, we obtain the following gradient of the O/H ratio: 
\begin{equation}
\mathrm{12+log(O/H)} = 8.84(\pm0.09) - 0.042(\pm0.009) \times R_{G},
\label{Eq:OH}
\end{equation}
\noindent with a dispersion around the fit of 0.07 dex. This observational result constrains the slope of the radial gradient of O until $R_G$ of 4 kpc using \ion{H}{ii} regions, providing a well-defined and apparently constant negative slope.  

Previous determinations of the radial gradient based on the additional sample made by \citet{Esteban:2018} reported an apparent break or flattening of the slope at $R_G \leq$ 8 kpc. Such result was reanalyzed by \citet{Arellano-Cordova:2020a}, who used revised Galactocentric distances based on $Gaia$ DR2 parallaxes, showing no evidence of that break in the inner Galactic disc, at least for $R_G \geq$ 6 kpc. The new observations included in this paper increase -- in a significant way -- the number of \ion{H}{ii} region with good-quality optical spectroscopical data in the inner disc, reaching well-determined $R_G$ as small as 4 kpc. We consider that Fig.~\ref{fig:O-gradient}  discards a possible break or flattening of the metallicity gradient at least in the range of $R_G$ covered by our new observations. The slope of our linear fit of the O gradient is in agreement with previous ones determined by our group in \citet{Esteban:2018} and \citet{Arellano-Cordova:2020a}, who find $-0.041$ and $-0.037$ dex kpc$^{-1}$, respectively, with a similar dispersion around the gradient of 0.07 dex and within of the observational errors, confirming that the gas is well mixed along the Galactic disc, giving no room for substantial azimuthal variations, at least in the quadrant of the disc covered by our complete sample of 42 Galactic H~{\sc ii} regions (see Fig.~\ref{fig:sample}). 

 Other previous optical studies covering a $R_G$ range between 6 and 18 kpc have reported slopes in the O/H gradient between $-0.028$ and $-0.061$ dex kpc$^{-1}$, which are roughly in agreement with our determinations, but with a significant higher dispersion around the gradient that can reach values up to 0.3 dex or even higher \citep{Shaver:1983, Vilchez:1996, Deharveng:2000, Rudolph:2006, Fernandez-Martin:2017}.
 
 The value of the dispersion with respect to the O/H gradient we report in this work is also consistent with the results obtained for other nearby disc galaxies with abundances calculated using direct determinations of the electron temperature \citep{ Bresolin:2011, Zurita:2012, Croxall:2016a, Esteban:2020, Berg:2020}, where no evidences of azimuthal variations were reported.

\subsection{The radial gradients of N/H and N/O}
\label{sec:nitrogen}

\begin{figure*} 
\begin{center}
    \includegraphics[width=0.9\textwidth, trim=30 0 30 0,  clip=yes]{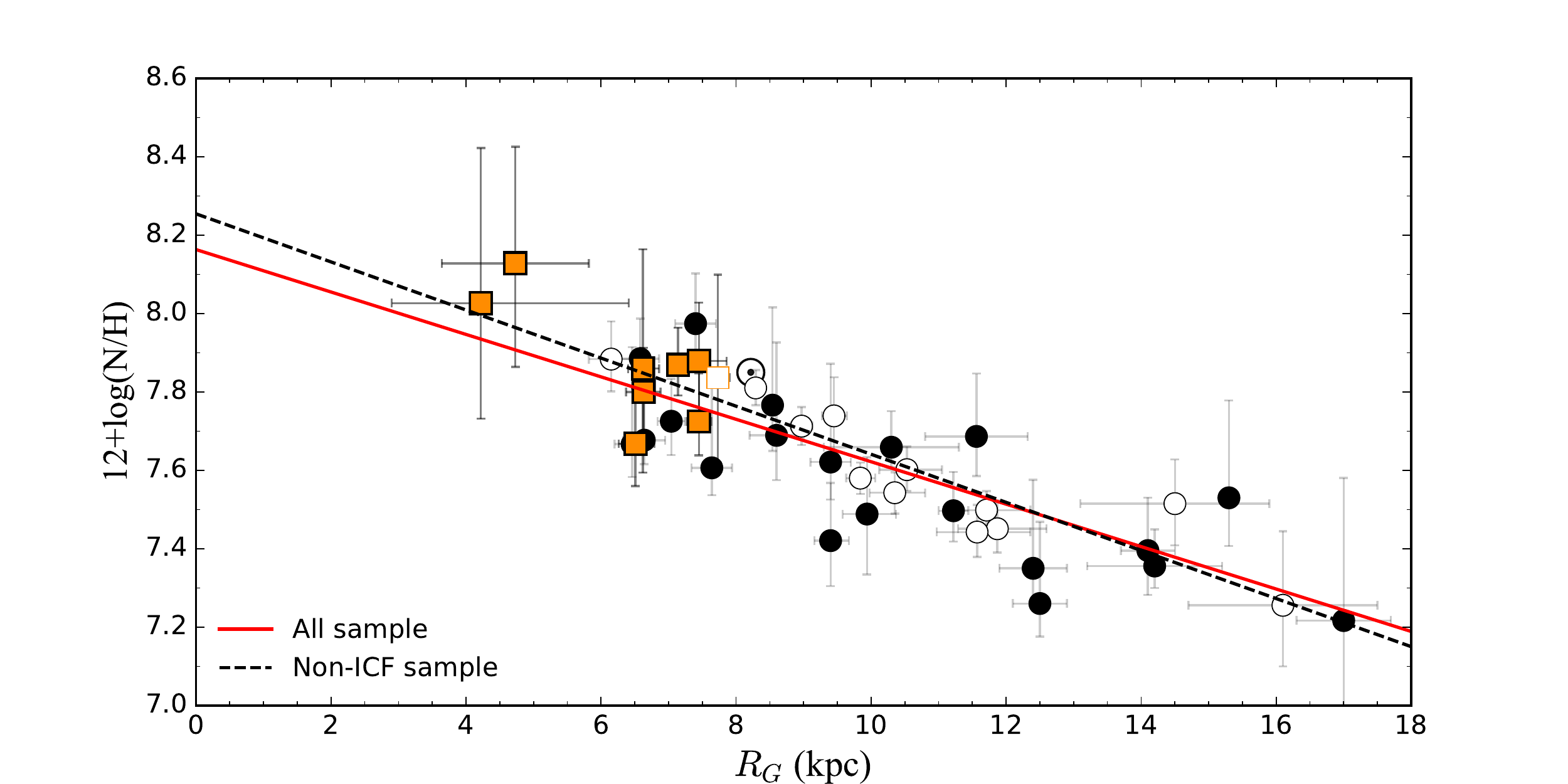}
    \caption{The radial abundance gradient of N for a sample of 41 Galactic \ion{H}{ii} regions as a function of the Galactocentric distance ($R_G$). The squares represents the data of the objects presented in this work and the circles the data of the additional sample of \ion{H}{ii} regions compiled by \citet{Arellano-Cordova:2020a}. The empty symbols show the objects whose N abundance was calculated without using an ICF scheme. The solid and dashed lines indicate the least-squares fit for all objects and only for those whose N abundance was calculated without an ICF, respectively. The solar symbol at 8.2 kpc indicates the photosphere N abundance of the Sun recommended by \citet{Lodders:2019}.}
    \label{fig:N-gradient}
   \end{center}
    \end{figure*}

Fig.~\ref{fig:N-gradient} shows the radial distribution of N abundance as a function of $R_G$ for a sample of 41 Galactic \ion{H}{ii} regions, that includes the nine new objects presented in this paper (squares) and 32 of the additional sample (circles). In this figure, we have discarded Sh\,2-212 -- of the additional sample -- owing to its odd N/O abundance ratio \citep[see][]{Arellano-Cordova:2020a}. The empty circles show those regions whose N abundances were calculated without the use of an ICF scheme. 

We compute the least-square fit for all the data represented in Fig.~\ref{fig:N-gradient} finding the following gradient:
\begin{equation}
\mathrm{12+log(N/H)} = 8.18(\pm0.11) - 0.057(\pm0.011) \times R_{G}.
\label{Eq:NH_1}
\end{equation}
\noindent and for the subsample of 13 regions with a non-ICF scheme -- empty symbols in Fig \ref{fig:N-gradient} -- as follows:
\begin{equation}
\mathrm{12+log(N/H)} = 8.24(\pm0.17) - 0.060(\pm0.017) \times R_{G}.
\label{Eq:NH_2}
\end{equation}
Fig.~\ref{fig:N-gradient} shows that the slope of the N/H gradient for the non-ICF sample is slightly steeper than that calculated for the whole sample, however, both slopes agree within the uncertainties. 
We obtain a dispersion around the fit of 0.11 dex for all the objects and 0.08 dex for the non-ICF subsample, respectively, again reinforcing the conclusion that the gas is well mixed along the Galactic disc. 

Very recently, we reported a slope of $-$0.049$\pm$0.006
 dex kpc$^{-1}$ for the N/H gradient considering the 32 Galactic \ion{H}{ii} regions of the additional sample  \citep{Arellano-Cordova:2020a} covering the range of $R_G$ between 6 and 17 kpc. In that work, we used the ICF(N$^{+}$/O$^+$) of \citet{Peimbert:1969}, which provides values of the N abundance quite similar to the ones obtained using the ICF scheme of Amayo et al. (in preparation) but more consistent with the positions of the non-ICF sample objects in the N/O $versus$ O/H diagram \citep[see Figure 4 of][]{Arellano-Cordova:2020a}. 

 We have compared the radial gradients obtained using N abundances calculated using both ICF schemes, Amayo et al. and \citet{Peimbert:1969}. In Fig. \ref{fig:N-gradient-PC69}, we show the results of the N/H $versus$ $R_G$ distribution and the N/H gradient we obtain when the abundances are calculated using the classical ICF scheme of N/O $\approx$ N$^{+}$/O$^{+}$ proposed by \citet{Peimbert:1969}. We obtain a radial gradient of 12+log(N/H) = $8.14(\pm0.07) -0.055(\pm0.007)\times R_G$, which is consistent with the one given in Eq. \ref{Eq:NH_1} based on the ICF scheme by Amayo et al. The data represented in \ref{fig:N-gradient-PC69} give a dispersion around the gradient line of 0.12 dex,  slightly larger than the one obtained with the ICF by Amayo et al. Moreover, the match between the empty symbols of Fig.~\ref{fig:N-gradient-PC69} -- which represent objects whose N/H ratio has been derived without an ICF -- with the N/H ratios determined with the ICF of \citet{Peimbert:1969} is somewhat poorer than when using the ICF by Amayo et al. (Fig.~\ref{fig:N-gradient}). These are the two reasons that make us to prefer using the ICF scheme of Amayo et al. for our final N abundances. One may note that error bars of the data are significantly smaller in Fig~\ref{fig:N-gradient-PC69} than in Fig.~\ref{fig:N-gradient}. This is an artifact because the ICF scheme \citet{Peimbert:1969} does not include any error criteria -- it is assumed to be zero -- while Amayo et al. provide numerical expressions to evaluate the uncertainties.  

\begin{figure*} 
\begin{center}
    \includegraphics[width=0.9\textwidth, trim=30 0 30 0,  clip=yes]{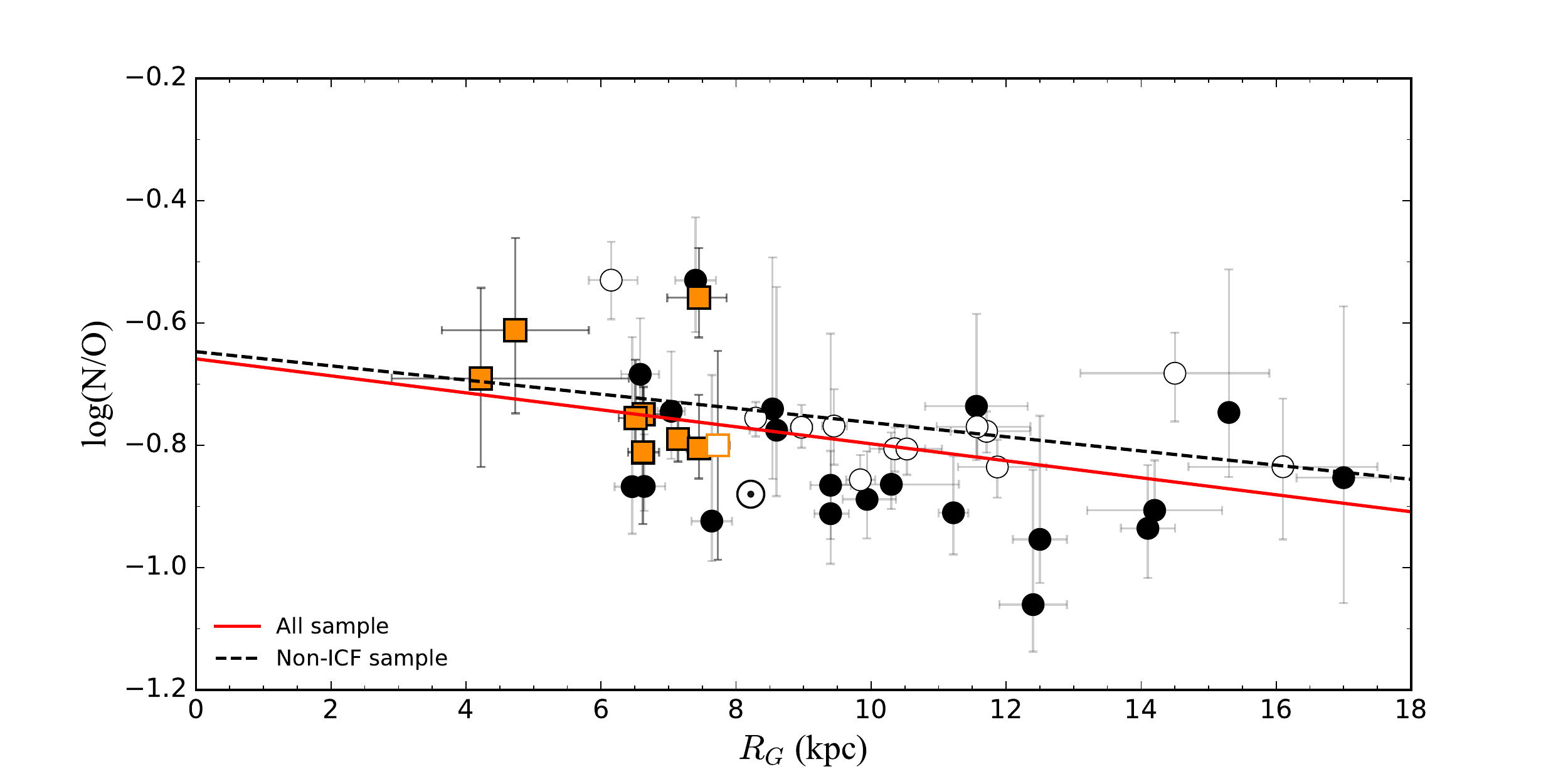}
    \caption{The radial gradient of N/O ratio for a sample of 41 Galactic \ion{H}{ii} regions as a function of the Galactocentric distance ($R_G$). The empty symbols show the objects whose N/H abundance was calculated without using an ICF scheme. The squares represents the data of the objects presented in this work and the circles the data of the additional sample of \ion{H}{ii} regions compiled by \citet{Arellano-Cordova:2020a}. The solid and dashed lines indicate the least-squares fit for all data and for those objects with non-ICF scheme, respectively. The solar symbol at 8.2 kpc indicates the photosphere N/O ratio of the Sun recommended by \citet{Lodders:2019}.}
    \label{fig:NO-gradient}
   \end{center}
    \end{figure*}

In Fig.~\ref{fig:NO-gradient}, we report the N/O gradient for a sample of 41 Galactic \ion{H}{ii} regions. The symbols are the same as in Fig~\ref{fig:N-gradient}. The linear least-square fit for the whole sample gives the following parameters:
\begin{equation}
\mathrm{log(N/O)} = -0.65(\pm0.07) - 0.015(\pm0.007) \times R_{G},
\label{Eq:NO_1}
\end{equation}
\noindent and for the sample of 13 \ion{H}{ii} regions whose N/H ratio has been calculated without and ICF, we obtain the following:
\begin{equation}
\mathrm{log(N/O)} = -0.65(\pm0.11) - 0.012(\pm0.010) \times R_{G}.
\label{Eq:NO_2}
\end{equation}
In Fig.~\ref{fig:NO-gradient}, we also represent the linear fits given in  Eqs.~\ref{Eq:NO_1} and ~\ref{Eq:NO_2} as solid and dashed lines, respectively.
The N/O gradients derived for both samples are very similar and also consistent with those reported by \citet{Esteban:2018} and \citet{Arellano-Cordova:2020a}. Eq.~\ref{Eq:NO_1} gives a rather flat slope, while  Eq.~\ref{Eq:NO_2} can be considered flat within the errors. The slight slope of the N/O gradient is produced by the relatively larger N/O ratios of three objects: Sh\,2-61, Sh\,2-90 and Sh\,2-93, located at $R_G$ between 6 and 8 kpc and showing log(N/O) between $-$0.56 and $-$0.53. The two last objects, Sh\,2-90 and Sh\,2-93 are at the same distance and located only 1.3$^\circ$ away, in the surroundings of the Vul OB1 Association. It is curious that those three objects are not located at abnormal positions in the radial O/H and N/H distributions shown in figs.~\ref{fig:O-gradient} and \ref{fig:N-gradient}. This suggest that their high N/O ratios may be perhaps a combination of slightly smaller O/H and larger N/H with respect to normal values. This behaviour is qualitatively consistent with the enrichment pattern observed in ring nebulae associated with Wolf-Rayet (WR) stars composed by stellar ejecta material \citep[e.g.][and references therein]{Esteban:2016}. One might speculate that a possible explanation of the chemical pattern of these three objects may be that they contain some chemically enriched material ejected in the past by a neighboring WR star of their vicinity. However, consulting the census of Galactic WR stars\footnote{http://pacrowther.staff.shef.ac.uk/WRcat/index.php} \citep{Rosslowe:2015}, we do not find a suitable candidate in the immediate vicinity of any of the three \ion{H}{ii} regions.
%
%
%
Excluding these three high-N/O objects from the total sample, the slope of the N/O gradient becomes almost flat, $-0.010\pm0.007$ dex kpc$^{-1}$. On the other hand, considering only the objects for which we do not use an ICF to derive their N/H ratio and discarding Sh\,2-61, the N/O ratio distribution may be represented by a constant value of log(N/O) = $-0.79\pm0.05$.


We have also determined the N/O gradient using the ICF scheme of \citet{Peimbert:1969} (Fig. \ref{fig:N-gradient-PC69}). The linear least-squares fit of the data provides a gradient of log(N/O) = $-0.67(\pm0.05) -0.016(\pm0.005)\times R_G$, fairly similar to the one obtained with the ICF scheme by Amayo et al. (Eq.~\ref{Eq:NO_1}). These data show a dispersion around the gradient of 0.11 dex, only slightly larger than the value of 0.10 dex obtained with the N/H determinations obtained with the ICF of Amayo et al.

\subsection{The N/O \textbf{\emph{versus}} O/H relationship in the Galaxy}
\label{sec:NO-OH}
\begin{figure} 
\begin{center}
    \includegraphics[width=0.45\textwidth, trim=30 0 30 0,  clip=yes]{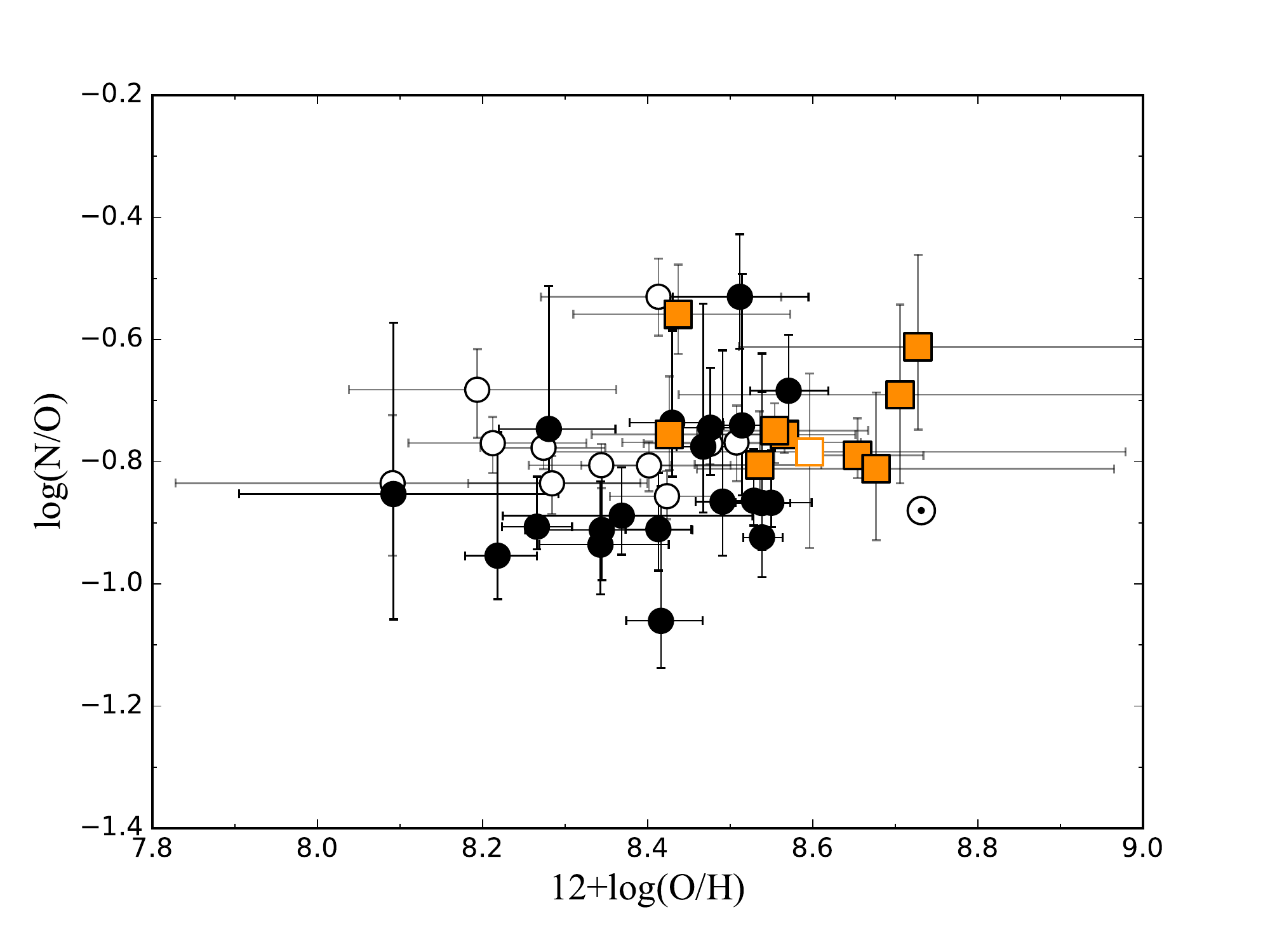}
    \caption{The N/O $versus$ O/H relation for a sample of 41 Galactic \ion{H}{ii} regions. The empty symbols show the objects where the N/H abundance was calculated without using an ICF scheme. The black squares represents the new data presented in this work and the red circles the data for the additional sample of \ion{H}{ii} regions compiled by \citet{Arellano-Cordova:2020a}. The solar symbol at 12+log(O/H) = 8.73 indicates the photosphere values of N/O and O/H ratios recommended by \citet{Lodders:2019}.}
    \label{fig:NO-OH}
   \end{center}
    \end{figure}

Owing to the different origins of N and O, the analysis of the N/O abundance ratio provides an important constraint about the chemical enrichment of our Galaxy. 
Fig \ref{fig:NO-OH} shows the well-known N/O $versus$ O/H diagram for our whole sample of 41 Galactic \ion{H}{ii} regions. Symbols are the same than in Fig.~\ref{fig:N-gradient}. 
 In Fig \ref{fig:NO-OH}, we can see that the data for the nine new objects presented in this work show 12+log(O/H) between 8.4 and 8.7, while their log(N/O) values are, in general, rather constant. We also note that the position of our new data in the N/O $versus$ O/H diagram is in agreement with that of the bulk of the additional sample. In the case of the objects whose N abundance was calculated without using an ICF scheme, we find a constant value of log(N/O) =  $-0.77\pm0.08$ ($-0.79\pm0.05$ excluding Sh\,2-61). As expected, the three aforementioned objects showing the highest N/O values, Sh\,2-61, Sh\,2-90 and Sh\,2-93 show log(N/O) values above the mean in Fig.~\ref{fig:NO-OH}.

\citet{Esteban:2020}, compiled high-quality observations (with direct determinations of $T_e$) \ion{H}{ii} regions in the Milky Way, M\,31, M\,101 and M\,33 and showed the N/O $versus$ O/H behaviour in these galaxies. They found that the N/O remains constant -- and shows log(N/O) $>$ $-$1.00 in practically all cases -- along the discs of the Milky Way and M\,31 for a range of 12+log(O/H) between 8.0 and 8.7. On the contrary, they found a rather different behavior in M\,33 and M\,101 (see their Figure 17). In fact, at such range of metallicities, we expect a clear increase of the N/O ratio with respect to O/H due to the secondary production of N by intermediate-mass stars \citep[e.g.][]{Berg:2019, Berg:2020, Esteban:2020}.

The flat N/O gradient and N/O $versus$ O/H distribution in the Galaxy indicate that the bulk of the N should not be formed 
by standard secondary processes along the Galactic disc. The  origin of this behaviour is difficult to understand. Possible explanations have been discussed in \citet{Esteban:2018, Esteban:2020} and \citet{Arellano-Cordova:2020a}.

\subsection{Comparison with results from infrared spectroscopy of \ion{H}{ii} regions}

Mid (3-8 $\mu$m) and especially FIR (15-1000 $\mu$m) spectroscopy of \ion{H}{ii} regions allows us to measure fine structure [\ion{C}{ii}], [\ion{N}{ii}], [\ion{N}{iii}], [\ion{O}{i}], [\ion{O}{iii}], [\ion{Ne}{ii}], [\ion{Ne}{iii}], [\ion{S}{iii}], [\ion{S}{iv}], [\ion{Ar}{iii}] and [\ion{Ar}{iv}] lines. The emissivity of those lines are less sensitive to $T_{\rm e}$ than optical ones, however, they may be more affected by $n_{\rm e}$ due to the low critical density of levels involved in the transitions. Most FIR spectroscopical studies assume an $ad\,hoc$ representative value of $T_{\rm e}$ \citep[][ although these last authors explore the effect of a Galactic $T_{\rm e}$ gradient on the results] {Rudolph:1997,Rudolph:2006,Martin-Hernandez+2002} or use values of $T_{\rm e}$ obtained from measurements of radio observations \citep{Shaver:1983, Fich:1991}, taken with instruments and  aperture sizes different than those of the FIR observations. 

Owing to FIR line intensities are virtually insensitive to dust extinction, they permit to determine chemical abundances in nebulae located at small Galactocentric distances, even reaching objects close to the Galactic centre. 

\begin{figure} 
\begin{center}
    \includegraphics[width=0.43\textwidth, trim=30 0 30 0,  clip=yes]{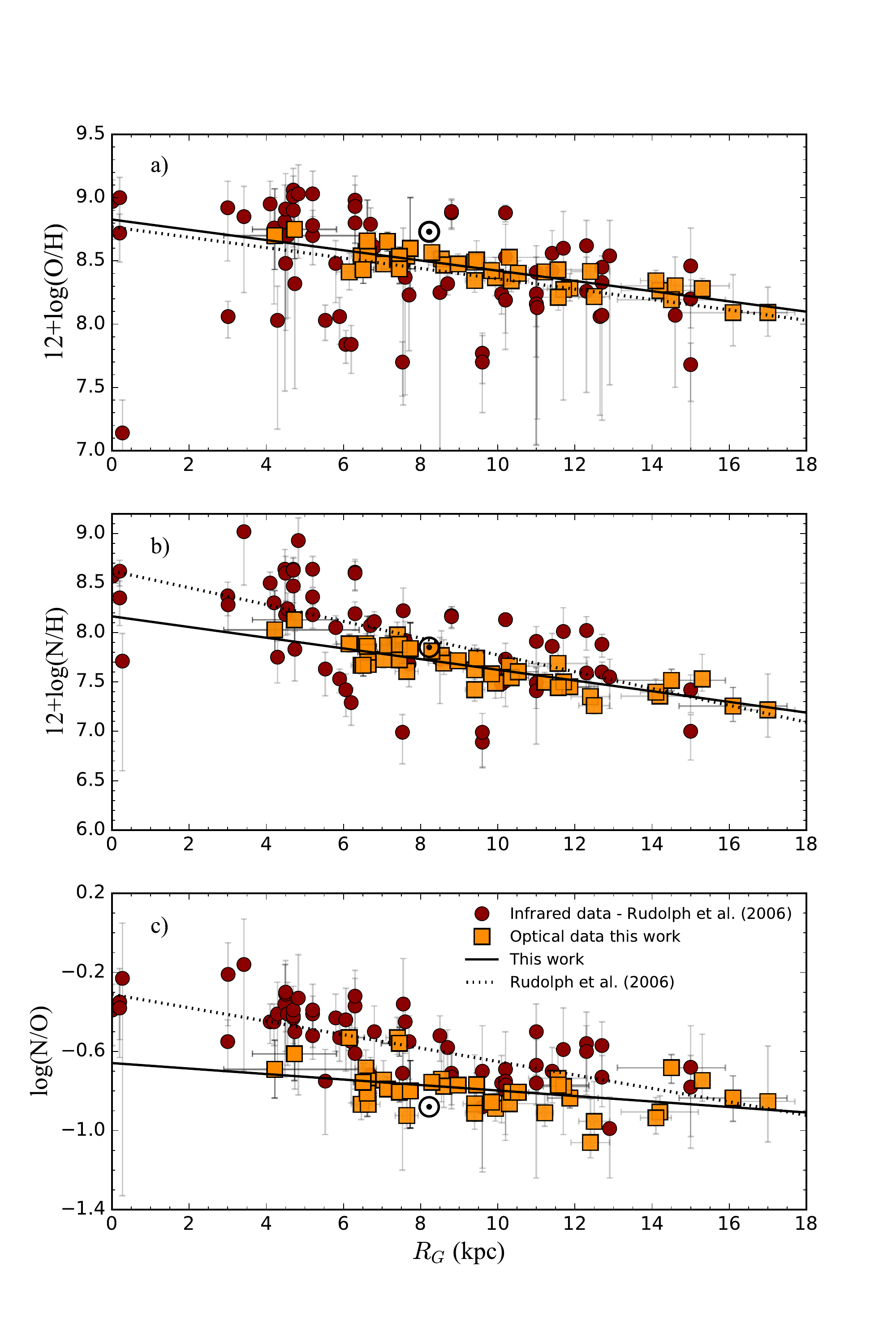}
    \caption{ From top to bottom: O/H, N/H, and N/O gradients for Galactic \ion{H}{ii} regions as a function of the Galactocentric distance ($R_G$). The squares represent our results based on optical data and the circles the results obtained by \citet{Rudolph:2006} from FIR observations. The solid line represents the fits obtained in this work and the dotted lines those calculated by \citet{Rudolph:2006}. Each plot shows the solar symbol at 8.2 kpc indicating the photosphere values of O/H, N/H, and N/O recommended by \citet{Lodders:2019}.}
    \label{fig:FIR-Comparison}
   \end{center}
    \end{figure}

Fig.~\ref{fig:FIR-Comparison} shows a comparison of the abundance gradients of O/H, N/H, and N/O using the results of the optical data (squares) obtained in this work and those by \citet{Rudolph:2006} from FIR data (circles). The solid and dotted lines indicate the value of the slopes for the optical and FIR data, respectively. 

\citet{Rudolph:2006} reported a slope of $-0.041\pm0.014$ dex kpc$^{-1}$ for O/H, which is in rather good agreement with our results using optical data (Eq.~\ref{Eq:OH}). However, panel (a) of Fig.~\ref{fig:FIR-Comparison} illustrates very clearly that the dispersion around the gradient is far much larger in the case of the FIR data. Unfortunately, this large dispersion gives inconclusive information about the exact slope of the O/H gradient in the innermost part of the Galaxy. We stress the importance of obtaining good quality data at such small $R_G$ to assess this problem.

For N/H, \citet{Rudolph:2006} reported a slope of $-0.085\pm0.010$ dex kpc$^{-1}$, which is significantly steeper than our slope of $-0.057\pm0.011$ (see Eq.~\ref{Eq:NH_1}). Panel (b) of Fig.~\ref{fig:FIR-Comparison} shows the comparison of the radial distribution of N/H abundances and its gradient using both optical and FIR data. We can see that the N/H ratios derived from FIR data tend to be systematically higher than the optical ones, especially for $R_G$ $\leq$ 6 kpc.  
Again, FIR results show a much larger dispersion around the fit in comparison with the optical sample of this study.

\citet{Rudolph:2006} reported a significant offset between optical and FIR data of the N/O ratios of Galactic \ion{H}{ii} regions, a fact that is also illustrated in panel (c) of Fig. \ref{fig:FIR-Comparison}. 
While the gradient based on FIR spectroscopy gives a clear negative slope of $-0.034\pm0.006$ dex kpc$^{-1}$ -- \citet{Martin-Hernandez+2002} report an even more negative slope of $-0.056\pm0.009$ dex kpc$^{-1}$ -- optical data give an almost flat one ($-0.015\pm0.007$ dex kpc$^{-1}$, Eq. \ref{Eq:NO_1}).
The reason for this so different behaviour is still unclear and is not easy to assess. First, in contrast to the optical calculations presented in this paper, the $T_{\rm e}$ used to calculate ionic abundances from FIR data is not determined from the same spectrum. However, this drawback of FIR observations alone cannot explain the discrepancy because of the low sensitivity to $T_{\rm e}$ of the emissivity of FIR lines. One more likely explanation could be the high dependence of the FIR fine-structure lines on $n_{\rm e}$ because of their low critical densities ($n_{\rm c}$). \citet{Martin-Hernandez+2002} showed that a correction factor is necessary to apply to derive ionic abundances when a low-density limit is assumed. In fact, when  $n_{\rm e}$ $>$ $n_{\rm c}$, the N$^{2+}$/H$^{+}$ ratio will be significantly underestimated by up a factor $\approx$ 2 or even larger, as reported for S$^{2+}$/H$^{+}$ and S$^{3+}$/H$^{+}$. Considering that [\ion{O}{iii}] and [\ion{N}{iii}] lines are produced in practically the same emitting zone of the nebulae, \citet{Martin-Hernandez+2002} argue that such problems with density may not be important for determining the N$^{2+}$/O$^{2+}$ ratio from FIR lines. However, this conclusion may be somewhat optimistic. The $n_{\rm e}$ values obtained from the ratio of [\ion{O}{iii}] $\lambda\lambda$52, 88$ \mu$m lines should correspond to an external layer, less denser than the inner core of the nebula that produces the bulk of the gas emission. One must take into account that \ion{H}{ii} regions observed with FIR spectroscopy tend to be compact or ultracompact objects, much denser and younger than optical \ion{H}{ii} regions. In fact, \citet{Martin-Hernandez+2002} also noted that the $n_{\rm e}$ obtained from  [\ion{O}{iii}] $\lambda\lambda$52, 88$ \mu$m lines are systematically lower than those obtained using radio observations in their sample of ultracompact \ion{H}{ii} regions. Some of the objects observed by those authors were also included in the compilation of \citet{Rudolph:2006}. 

We consider that the ICF scheme is the most important potential source of uncertainty in the estimation of the N/O ratio from FIR data. To try to assess this possibility, we have analysed the data presented by \citet{Rudolph:2006}, which include their own observations for some objects and data compiled from previous works by \citet{Simpson:1995, Afflerbach:1997, Rudolph:1997} and \citet{Peeters:2002} (analysed in  \citealt{Martin-Hernandez+2002}). We have taken information about $R_G$, ionic and total abundances of O and N, and the different results of their ICF schemes.
In a subsample of 12 objects of the data compiled by \citet{Rudolph:2006} is also possible to calculate the N$^{+}$/H$^+$ ratio using the measurement of the [\ion{N}{ii}] $\lambda$121.7 $\mu$m line published by  \citet{Martin-Hernandez+2002}. This allows to calculate N/H by simply adding the contribution of N$^{+}$/H$^{+}$ and N$^{2+}$/H$^{+}$, without the use of an ICF scheme for those objects \citep{Martin-Hernandez+2002}. However, due to the lack of [\ion{O}{ii}] lines in the FIR range, the use of an ICF(N$^{2+}$/O$^{2+}$) scheme is still necessary to calculate the N/O abundance \citep{Rubin:1988, Martin-Hernandez+2002, Rudolph:2006}. It must be taken into account that, at smaller $R_G$, N$^{+}$ and O$^{+}$ tend to be the dominant ions of N and O, and larger corrections to the observed N$^{2+}$/O$^{2+}$ ratios are necessary to estimate the N/O ratio for objects at the inner Galactic disc. For instance, using photoionization models,  \citet{Martin-Hernandez+2002} showed that the values of the ICF(O$^{2+}$) for those inner objects are significantly larger, with a strong dependence on the ionization degree.

\begin{figure} 
\begin{center}
    \includegraphics[width=0.35\textwidth, trim=30 0 30 0,  clip=yes]{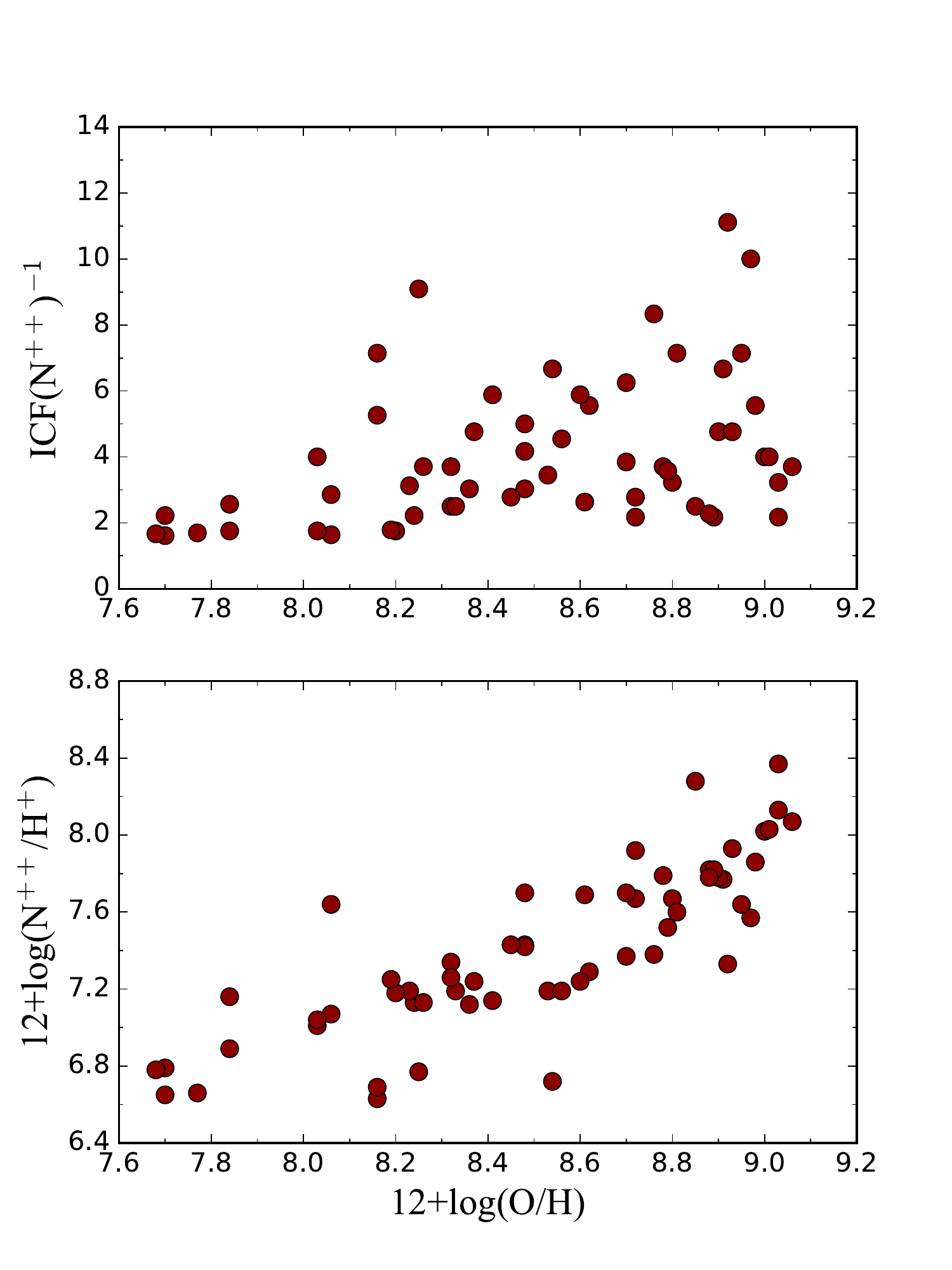}
    \caption{Data obtained from FIR spectroscopy compiled by \citet{Rudolph:2006} for a sample of Galactic \ion{H}{ii} regions. Upper panel: ICF(N$^{2+}$) $versus$ 12+log(O/H). Bottom panel: 12+log(N$^{2+}$/H$^{+}$) $versus$ 12+log(O/H).}
    \label{fig:ICF-FIR}
   \end{center}
    \end{figure}

In the upper and bottom panels of Fig.~\ref{fig:ICF-FIR}, we present the values of the ICF(N$^{2+}$) and N$^{2+}$/H$^{+}$ ratio as a function of the O abundance taken from the results compiled by \citet{Rudolph:2006}. The upper panel of Fig.~\ref{fig:ICF-FIR} illustrates that the ICF(N$^{2+}$) --based on photoionization models -- shows some dependence on O/H, being the  correction larger at higher O/H values. On the other hand, the bottom panel of Fig.~\ref{fig:ICF-FIR} shows a clear correlation between  N$^{2+}$/H$^+$ and O/H ratios. This correlation seems rather artificial because the N$^{2+}$/H$^+$ of the nebulae should depend basically on two factors: the metallicity and the ionization degree. One would expect a much dispersed and less defined correlation in the distribution of those two quantities. From the behavior we see in both panels of Fig.~\ref{fig:ICF-FIR}, one cannot rule out that the N/H ratio determined by \citet{Rudolph:2006} may be overestimated at high metallicities (small $R_G$).

 The use of different atomic data and methodologies to calculate the physical conditions and chemical abundances may also introduce part of the difference between the results obtained from optical and FIR data. \citet{JuandeDios:2017} showed that changes in the atomic data can lead to differences in the O/H, N/H, and N/O ratios of up to 0.2 dex for regions with $n_{\rm e} \leq 10^{3}$ cm$^{-3}$, a difference that can easily increase at higher densities. In fact, the densities computed from FIR data for \ion{H}{ii} regions in the inner part of the Galactic disc ($R_G < 8.2$ kpc) can reach values up to 8400 cm$^{-3}$ \citep{Rudolph:2006}. Additionally, \citet{JuandeDios:2017} also noted that the N/O abundance ratio is sensitive to changes in atomic data.
In this regard, it is noteworthy that the large scatter seen in the O/H and N/H gradients from FIR data seen in Fig.~\ref{fig:FIR-Comparison} is reduced considerably in the N/O gradient. This effect can be owing to the high sensitivity to density of FIR O/H and N/H ratio determinations, since FIR fine-structure line emissivities are strongly dependent on the relative populations between the fine-structure levels. This dependence is somewhat diluted when computing the N/O ratio as the same density for both abundance determinations is applied, making observational errors and uncertainties associated to the adopted ICF the main contributors to the observed N/O scatter.

\begin{figure} 
\begin{center}
    \includegraphics[width=0.42\textwidth, trim=30 0 30 0,  clip=yes]{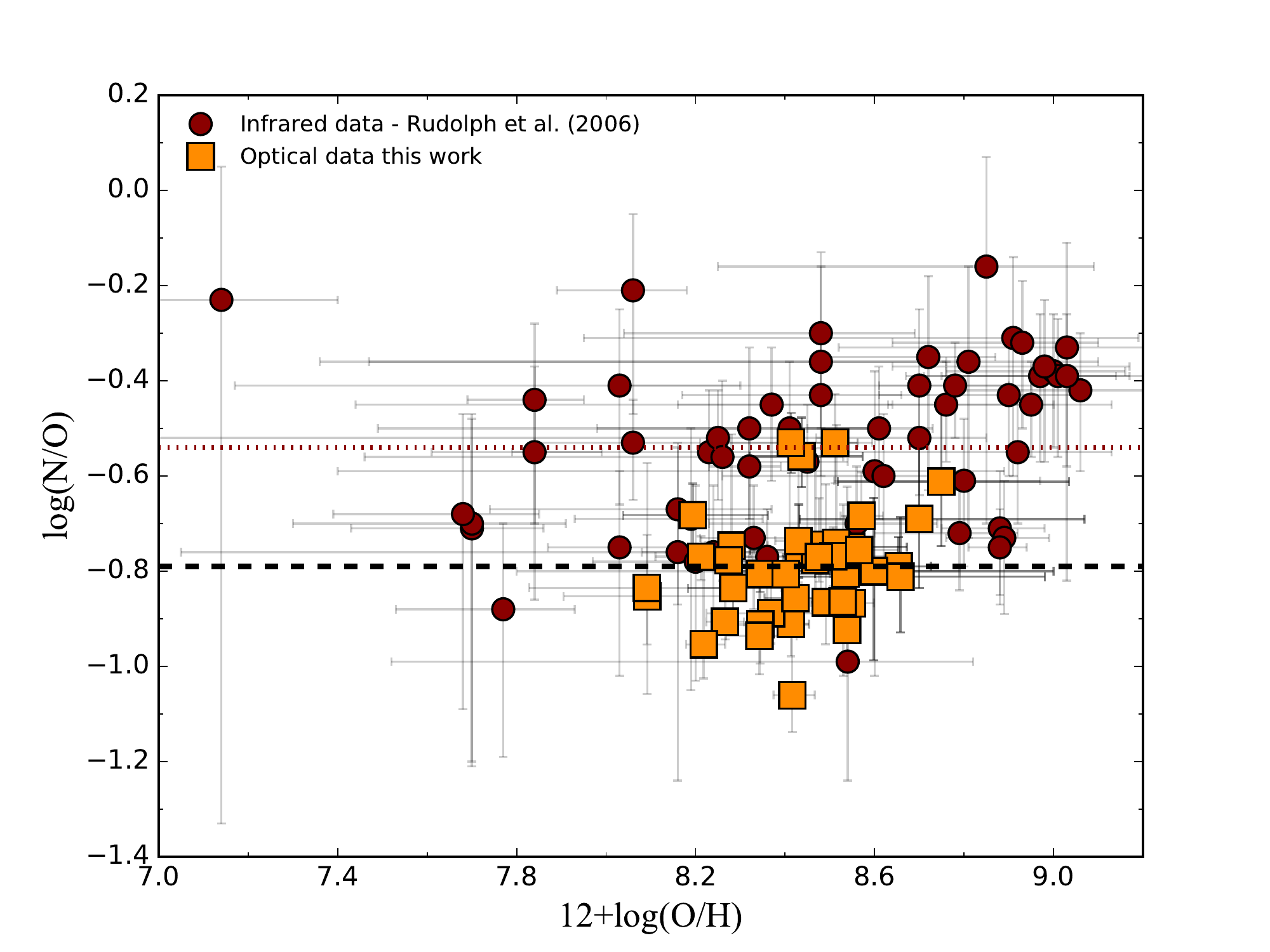}
    \caption{The N/O $versus$ O/H relation for a sample of Galactic \ion{H}{ii} regions from optical and FIR observations. The squares identify the optical data from this work and the circles the FIR data compiled by \citet{Rudolph:2006}. The dotted and dashed lines are the mean value of log(N/O) for the FIR and optical data, respectively.}
    \label{fig:ICF-FIR-OPT-NO}
   \end{center}
    \end{figure}

We have also compared the FIR and optical samples in the N/O $versus$ O/H diagram. Fig.~\ref{fig:ICF-FIR-OPT-NO} shows the ratter different behaviour of these two sets of data. The dashed and dotted lines indicate the mean values of log(N/O) of the optical and FIR observations, respectively. FIR data show a much larger dispersion and tend to give higher values of N/O with a mean value of log(N/O) = $-$0.54 $\pm$ 0.18 dex. As it was discussed in Section~\ref{sec:NO-OH} in the case of optical data, the distribution of N/O ratios as a function of O/H determined from FIR observations does not show a clear trend of increasing N/O values at high metallicities.  

Summarizing, considering the discussion above, the disagreement of the N/O ratios obtained from FIR and optical data -- which is especially remarkable at small $R_G$ values -- may be an artifact of the methodology to derive abundances from FIR observations due to the combination of these two factors: (a) the strong dependence of the N/O ratio on the adopted ICFs and (b) the possible effects of $n_{\rm e}$ on the calculation of ionic abundances. The chemical abundances determined from optical data presented in this paper are comparatively more robust. Firstly, they are based on direct determinations of $T_{\rm e}$ obtained from the same spectrum. Secondly, the $n_{\rm e}$ values used are representative of the whole nebula and the low-density limit is applicable for most objects. Thirdly, the O/H ratio is determined without the use of an ICF. Finally -- and this is important for objects located at small $R_G$ values, where the optical-FIR disagreement is larger -- the bulk of N in low-ionization degree nebulae -- the dominant population of \ion{H}{ii} regions at small $R_G$ and high metallicities -- is in the form of N$^+$, which abundance can be determined from optical [\ion{N}{ii}] lines. Therefore, the impact of IFC(N$^+$/O$^+$) onto N/O determinations based on optical data is much less important than that of IFC(N$^{2+}$/O$^{2+}$) onto FIR-based  determinations. Moreover, the different ICFs used for optical data give similar N/H values and some of them \citep[e.g.][]{Peimbert:1969} have been widely tested \citep[see discussion in][]{Arellano-Cordova:2020a}.

\section{Conclusions}
\label{sec:conclusions}
We present spectra of nine \ion{H}{ii} regions observed with the 10.4 m GTC telescope, with $R_G$ from 4 to 8 kpc. Their distances have been revised using $Gaia$ parallaxes of the second data release, DR2 \citep{Bailer-Jones:2018, gaiadr2}. We have direct determinations of $T_{\rm e}$ for all objects, which allows a precise calculation of their ionic abundances. We present ionic and total abundances of different elements, focusing on the analysis of the N and O abundances in order to constrain their radial abundance gradients at small $R_G$. In addition, we have included the results of the sample of Galactic \ion{H}{ii} regions compiled and analyzed by \cite{Arellano-Cordova:2020a}. Our final sample comprises spectra of 42 Galactic \ion{H}{ii} regions covering $R_G$ values from 4 to 17 kpc. 

We report linear radial abundance gradients for O/H and N/H with slopes of $-0.042\pm0.009$ dex kpc$^{-1}$ and $-0.057\pm0.011$ dex kpc$^{-1}$, respectively. The  results discard the presence of a possible flattening of the O/H gradient in the inner part of the Galactic disc as suggested by \citet{Esteban:2018}, at least for $R_G \geq$ 4 kpc. The dispersion of the O/H and N/H values with respect to the computed gradients is 0.07 and 0.11 dex, respectively, which implies that the gas through the Galactic disc is well-mixed. This result also indicates the absence of significant azimuthal variations of the chemical abundances, at least in the quadrant covered by our sample objects.

The radial gradient of N/O ratio shows a small negative slope of $-0.015\pm0.007$ dex kpc$^{-1}$, flatter than some other previous determinations \citep[e.g.][]{Martin-Hernandez+2002, Rudolph:2006}. This result is also reflected in an almost flat N/O $versus$ O/H diagram relation, a behavior that is not observed in other nearby spiral galaxies except M31 \citep{Esteban:2020}. We compare our results with those obtained from FIR data \citep[the compilation by][]{Rudolph:2006}. We find that our optical O/H and N/H data show significant less dispersion around the gradient lines than FIR data. Finally, we confirm the significant offset in the N/O distribution between the optical and FIR observations, especially at small values of $R_G$. A likely explanation for this discrepancy might be related to the ICF used in FIR studies to correct for the contribution of unseen N$^{+}$ and O$^+$ to the total abundances of N and O, ions that become dominant at small $R_G$. Another possibility is the strong dependence of the ionic abundances determined from FIR lines on the assumed value of  
$n_{\rm e}$.

\section*{DATA AVAILABILITY}
We provide the line intensities measurements of the spectra in Appendix~\ref{appendixA}.

\section*{Acknowledgements}
 We acknowledge the anonymous referee for his/her comments which have helped to improve the quality of the paper. We thank A. Amayo, G. Delgado-Inglada, G. and  Stasi\'nska for providing us their ICF computations prior to publication.
 KZA-C acknowledges support from Mexican CONACYT postdoctoral grant 364239.  We acknowledge support from the Agencia Estatal de Investigaci\'on del Ministerio de Ciencia e Innovaci\'on (AEI-MCINN) under grant {\it Espectroscop\'\i a de campo integral de regiones \ion{H}{ii} locales. Modelos para el estudio de regiones \ion{H}{ii} extragal\'acticas} with reference 10.13039/501100011033. JG-R acknowledges support from an Advanced Fellowship from the Severo Ochoa excellence program (SEV-2015-0548).  JEM-D thanks the support of the Instituto de Astrof\'isica de Canarias under the Astrophysicist Resident Program and acknowledges support from the Mexican CONACyT (grant CVU 602402). The authors acknowledge support under grant P/308614 financed by funds transferred from the Spanish Ministry of Science, Innovation and Universities, charged to the General State Budgets and with funds transferred from the General Budgets of the Autonomous Community of the Canary Islands by the MCIU.  This paper is based on observations made with the Gran Telescopio Canarias (GTC), installed in the Spanish Observatorio del Roque de los Muchachos of the Instituto de Astrof\'isica de Canarias, in the island of La Palma, Spain.
 This work has made use of data from the European Space Agency (ESA) mission
{\it Gaia} (\url{https://www.cosmos.esa.int/gaia}), processed by the {\it Gaia}
Data Processing and Analysis Consortium (DPAC,
\url{https://www.cosmos.esa.int/web/gaia/dpac/consortium}). Funding for the DPAC
has been provided by national institutions, in particular the institutions
participating in the {\it Gaia} Multilateral Agreement.




\bibliographystyle{mnras}
\bibliography{refs-MW} 


\appendix
\section{Line intensity ratios}
\label{appendixA}
In this Appendix, we present the line intensity ratios of the nine Galactic \ion{H}{ii} regions observed. The first two columns list the laboratory wavelength, and the ion identification.Column 3 identifies the multiplets to which each CEL or RL belong, following the classical nomenclature established by \citet{moore45},  and the identification of the corresponding line in hydrogen Balmer series lines. Column 4 gives the adopted reddening function, $f$($\lambda$). The remaining columns indicate the dereddened line intensity ratios with respect to H$\beta$ and the associated uncertainties. The last two rows of each table include the reddening coefficient and the observed -- uncorrected for reddening -- $F$(H$\beta$) of the extracted aperture for each object.
 

\onecolumn
\small
	\begin{longtable}{ccccc c c c c  }
			\caption{Derredened line intensiy ratios with respect to $I$(H$\beta$)=100 for Sh 2-29, Sh 2-32, Sh 2-47,  Sh 2-48 and Sh 2-53.} \label{TableA1}
\tabularnewline
\hline

 \multicolumn{1}{c}{$\lambda_{\rm 0}$} &     &     &    &   &   &  \multicolumn{1}{c}{$I(\lambda)$/$I$(H$\beta$)} \\
 \multicolumn{1}{c}{\AA}  &   \multicolumn{1}{c}{Ion}  & \multicolumn{1}{c}{ID}   & \multicolumn{1}{c}{$f$($\lambda$)}   & \multicolumn{1}{c}{Sh\,2-29} & \multicolumn{1}{c}{Sh\,2-32} & \multicolumn{1}{c}{Sh\,2-47} 
  & \multicolumn{1}{c}{Sh\,2-48} & \multicolumn{1}{c}{Sh\,2-53} \\   
\hline
\endfirsthead
\caption{continued.}\tabularnewline
\hline
 \multicolumn{1}{c}{$\lambda_{\rm 0}$}  &   \multicolumn{1}{c}{Ion}  & \multicolumn{1}{c}{ID}   & \multicolumn{1}{c}{$f$($\lambda$)}   & \multicolumn{1}{c}{Sh\,2-29} & \multicolumn{1}{c}{Sh\,2-32} & \multicolumn{1}{c}{Sh\,2-47}
 & \multicolumn{1}{c}{Sh\,2-48} & \multicolumn{1}{c}{Sh\,2-53} \\   

\hline
\endhead
\hline
\endfoot
\endlastfoot

3726	&	[\ion{O}{ii}]	      &	    1F	        &	          0.257	&  280.8$\pm$7.4     &   175.1$\pm$5.7      &    180.7$\pm$12.5    &     129.2$\pm$11.4       &   208.0$\pm$46.8   \\
3729	&	[{\ion{O}{ii}}]	      &	    1F	        &	                &   	 			 &                      &                      &                          &   $-$              \\
3750	&	{\ion{H}{i}}	      &	    H12	        &	          0.253	&  $-$               &   $-$                &   $-$                &      $-$                 &   $-$              \\
3771	&	{\ion{H}{i}}	      &	    H11	        &	          0.249	&  3.03$\pm$0.55     &   2.14$\pm$0.54      &   $-$                &      $-$                 &   $-$              \\
3798	&	{\ion{H}{i}}	      &	    H10	        &	          0.244	&  5.10$\pm$0.62     &   2.19$\pm$0.48      &   $-$                &      $-$                 &   $-$              \\
3835	&	{\ion{H}{i}}	      &	    H9	        &	          0.237	&  7.01$\pm$0.72     &   9.2$\pm$1.28       &   $-$                &      $-$                 &   $-$              \\
3869	&	[{\ion{Ne}{iii}}] 	  &	    1F	        &             0.211 &                    &                      &                      &                          &                    \\
3889	&	{\ion{He}{i}}	      &	    5	        &	         0.227	&  14.02$\pm0.94$    &   8.93$\pm$0.69      &   4.14$\pm$1.67      &      15.3$\pm$3.1        &   $-$              \\
3889	&	{\ion{H}{i}}	      &	    H8	        &	                &                    &                      &                      &                          &                    \\
3967	&	[{\ion{Ne}{iii}}] 	  &	    1F	        &             0.211 &  $-$               &   $-$                &   8.39$\pm$1.66      &      13.1$\pm$2.5        &   $-$              \\
3970    &   {\ion{H}{i}}	      &	    H7	        &	                &                    &                      &                      &                          &                    \\
4026	&	{\ion{He}{i}}	      &	    18	        &	          0.198	&  0.86$\pm$0.18     &   $-$                &   $-$                &      $-$                 &   $-$              \\
4069	&	[\ion{S}{ii}]	      &	    1F	        &	          0.189	&  2.65$\pm$0.27     &   2.07$\pm$0.42      &   $-$                &      $-$                 &   $-$              \\
4076	&	[\ion{S}{ii}]	      &	    1F	        &	          0.187	&  1.52$\pm$0.23     &   1.08$\pm$0.29      &   $-$                &      $-$                 &   $-$              \\
4102	&	{\ion{H}{i}}	      &	    H6	        &	          0.182	&  26.02$\pm$0.66    &   24.27$\pm$0.74     &   22.9$\pm$1.4       &      23.4$\pm$2.0        &   $-$              \\
4144	&	{\ion{He}{i}}	      &	    53	        &	          0.172	&  $-$               &   $-$                &   $-$                &      $-$                 &   $-$              \\
4267	&	{\ion{C}{ii}}	      &	    6	        &	          0.144	&  $-$               &   $-$                &   $-$                &      $-$                 &   $-$              \\
4340	&	{\ion{H}{i}}	      &	    H$\gamma$    &	          0.127	&  45.75$\pm$0.97    &   44.88$\pm$1.03     &   41.7$\pm$2.0       &      43.1$\pm$1.3        &   26.0$\pm$5.8     \\
4471	&	{\ion{He}{i}}	      &	    14	         &	          0.096	&  $-$               &   $-$                &   $-$                &      2.9$\pm$0.5         &   $-$              \\
4651	&	{\ion{O}{ii}}	      &	    1	        &	             	&                    &                      &                      &                          &                    \\
4861	&	{\ion{H}{i}}	      &	    H$\beta$	 &	          0.000	&  100.0$\pm$2.0     &   100.0$\pm$2.0      &   100.0$\pm$2.3      &      100.0$\pm$2.1       &   100.0$\pm$2.4    \\
4922	&	{\ion{He}{i}}	      &	    48	         &	       $-$0.015	&  0.49 $\pm$0.10	 &   $-$                &   $-$                &      0.8$\pm$0.2         &   $-$              \\
4959	&	[\ion{O}{iii}]	      &	    1F	         &	       $-$0.024	&  3.58 $\pm$0.12    &   $-$                &   3.0$\pm$0.4        &      9.4$\pm$0.5         &   $-$              \\
4986	&	[\ion{Fe}{iii}]       &	    3F	         &	       $-$0.031	&  0.15 $\pm$0.08    &   $-$                &   $-$                &      $-$                 &   $-$              \\
5007	&	[\ion{O}{iii}]	      &	    1F	         &	       $-$0.036	&  11.83$\pm$0.33    &   2.14$\pm$0.25      &   12.8$\pm$0.7       &      31.0$\pm$0.7        &   16.2$\pm$1.6     \\
5016	&	{\ion{He}{i}}	      &	    4   	     &         $-$0.038	&  0.84$\pm$0.10     &   $-$                &   $-$                &      1.0$\pm$0.2         &   $-$              \\
5198	&	[\ion{N}{i}]	      &	    1F	         &	       $-$0.082	&  1.49$\pm$0.12     &   2.26$\pm$0.19      &   1.4$\pm$0.2        &      0.5$\pm$0.1         &   $-$              \\
5270	&	[\ion{Fe}{iii}]	      &	    1F	         &	       $-$0.100	&  0.11$\pm$0.05     &   $-$                &   $-$                &      $-$                 &   $-$              \\
5518	&	[\ion{Cl}{iii}]       &	    1F	         &	       $-$0.154	&  0.16$\pm$0.05     &   $-$                &   $-$                &      $-$                 &   $-$              \\
5538	&	[\ion{Cl}{iii}]       &	    1F	         &	       $-$0.158	&  0.13$\pm$0.04     &   $-$                &   $-$                &      $-$                 &   $-$              \\
5755	&	[\ion{N}{ii}]	      &	    3F	         &	       $-$0.194	&  0.93$\pm$0.07     &   0.59$\pm$0.06      &   0.51$\pm$0.12      &      0.39$\pm$0.10       &   0.95$\pm$0.22    \\
5876	&	{\ion{He}{i}}	      &	    11	         &	       $-$0.215	&  5.02$\pm$0.16     &   $-$                &   1.3$\pm$0.2        &      13.5$\pm$0.5        &   6.9$\pm$0.9      \\
6300	&	[\ion{O}{i}]	      &   	1F	         &	       $-$0.282	&  3.48$\pm$0.26     &   $-$                &   $-$                &      $-$                 &   $-$              \\
6312	&	[\ion{S}{iii}]	      &   	3F	         &	       $-$0.283	&  0.28$\pm$0.10     &   $-$                &   $-$                &      $-$                 &   $-$              \\
6548    &   [\ion{N}{ii}]          &    1F           &         $-$0.318 &  47.9$\pm$1.2      &   36.1$\pm$0.01      &   33.9$\pm$2.0       &      28.1$\pm$1.3        &   63.2$\pm$2.1     \\
6563    &   {\ion{H}{i}}          &     H$\alpha$    &         $-$0.320 &  293.4$\pm$7.0     &   282.1$\pm$8.4      &   266.3$\pm$15.2     &      276.7$\pm$12.4      &   294.4$\pm$9.7    \\
6583    &   [\ion{N}{ii}]         &     IF           &         $-$0.323 &  147.4$\pm$3.6     &   109.9$\pm$3.2      &   102.1$\pm$6.0      &      83.7$\pm$3.8        &   187.0$\pm$6.2    \\
6678    &   {\ion{He}{i}}         &     46           &         $-$0.336 &  1.42$\pm$0.11     &   $-$                &   0.37$\pm$0.08      &      3.0$\pm$0.2         &   1.7$\pm$0.1      \\
6716    &   [\ion{S}{ii}]         &     2F           &         $-$0.342 &  56.5$\pm$1.4      &  46.93$\pm$1.46      &   46.4$\pm$3.0       &      22.8$\pm$1.1        &   25.2$\pm$0.9     \\
6731    &   [\ion{S}{ii}]         &     2F           &         $-$0.344 &  41.8$\pm$1.0      &  33.70$\pm$1.05      &   32.8$\pm$2.1       &      15.9$\pm$0.8        &   23.1$\pm$0.8     \\
7065	&	{\ion{He}{i}}	      &	    10     	     &         $-$0.387	&  5.25$\pm$0.08     &   $-$                &   1.1$\pm$0.2        &      1.1$\pm$0.1         &   0.82$\pm$0.07    \\
7136    &   [\ion{Ar}{iii}]       &     1F           &         $-$0.396 &  2.43$\pm$0.12     &   0.33$\pm$0.07      &   0.3$\pm$0.1        &      4.2$\pm$0.2         &   2.4$\pm$0.05     \\

\hline
$c$(H$\beta$)     &                &                 &                   &  0.46$\pm$0.02     & 0.52$\pm$0.03       &  1.07$\pm$0.07        &     1.53$\pm$0.05       &   3.25$\pm$0.04    \\                       
$F$(H$\beta$)$^{\rm a}$     &                &                 &                   &    4.8$\pm$0.1   &   3.54$\pm$0.07     &     0.97$\pm$0.02  &      0.85$\pm$0.02 &       0.249$\pm$0.006
\\
\hline
\multicolumn{9}{l}{$^{\rm a}$ Flux uncorrected for reddening in units of 10$^{-14}$ erg cm$^{-2}$ s$^{-1}$.}
\end{longtable}
\onecolumn
\small
	\begin{longtable}{ccccc c c c c}
			\caption{Derredened line intensiy ratios with respect to $I$(H$\beta$)=100 for Sh 2-54, Sh 2-82, Sh 2-88B and Sh 2-93.} \label{TableA2}
\tabularnewline
\hline
 \multicolumn{1}{c}{$\lambda_{\rm 0}$} &     &     &    &   &     \multicolumn{2}{c}{$I(\lambda)$/$I$(H$\beta$)} \\
 \multicolumn{1}{c}{\AA}  &   \multicolumn{1}{c}{Ion}  & \multicolumn{1}{c}{ID}   & \multicolumn{1}{c}{$f$($\lambda$)}  & \multicolumn{1}{c}{Sh\,2-54} & \multicolumn{1}{c}{Sh\,2-82} & \multicolumn{1}{c}{Sh\,2-88B} & \multicolumn{1}{c}{Sh\,2-93}\\   
\hline
\endfirsthead
\caption{continued.}\tabularnewline
\hline
 \multicolumn{1}{c}{$\lambda_{\rm 0}$}  &   \multicolumn{1}{c}{Ion}  & \multicolumn{1}{c}{ID}   & \multicolumn{1}{c}{$f$($\lambda$)}  & \multicolumn{1}{c}{Sh\,2-54}   & \multicolumn{1}{c}{Sh\,2-82} & \multicolumn{1}{c}{Sh\,2-88B} & \multicolumn{1}{c}{Sh\,2-93}\\   

\hline
\endhead
\hline
\endfoot
\endlastfoot

3726	&	[\ion{O}{ii}]	      &	    1F	       &             0.257	       &    140.1$\pm$7.9      &  179.2$\pm$18.0         &       282.6$\pm$11.9    &      211.7$\pm$17.5  \\   
3729	&	[{\ion{O}{ii}}]	      &	    1F	       &                          &                       &                         &       $-$                &                      \\   
3750	&	{\ion{H}{i}}	      &	    H12	       &             0.253	       &    1.9$\pm$0.8        &       $-$               &       $-$               &      $-$             \\   
3771	&	{\ion{H}{i}}	      &	    H11	       &             0.249	       &    3.4$\pm$1.2        &       $-$               &       $-$               &      $-$             \\   
3798	&	{\ion{H}{i}}	      &	    H10	       &             0.244	       &    5.0$\pm$1.2        &       $-$               &       3.1$\pm$1.2       &      $-$             \\   
3835	&	{\ion{H}{i}}	      &	    H9	       &             0.237	       &    6.6$\pm$1.0        &       $-$               &       5.2$\pm$2.1       &      $-$             \\   
3867	&	{\ion{He}{i}}	      &	    20	       &             0.231	       &    1.7$\pm$0.7        &       $-$               &       $-$               &      $-$             \\   
3869	&	[{\ion{Ne}{iii}}] 	  &	    1F	       &             0.211        &                       &                         &                          &                      \\   
3889	&	{\ion{He}{i}}	      &	    5	       &            0.227	       &    17.3$\pm$0.9       &       $-$               &       15.3$\pm$1.6      &      16.9$\pm$2.6    \\   
3889	&	{\ion{H}{i}}	      &	    H8	       &                          &                       &                      &                             &                      \\   
3967	&	[{\ion{Ne}{iii}}] 	  &	    1F	       &             0.211        &    16.6$\pm$1.2       &       $-$               &       14.8$\pm$1.1       &      25.5$\pm$3.4    \\   
3970    &   {\ion{H}{i}}	      &	    H7	       &                          &                       &                         &                          &      $-$             \\   
4009	&	{\ion{He}{i}}	      &	    55	       &             0.202	       &    3.2$\pm$0.6        &       $-$               &       $-$               &      $-$             \\   
4026	&	{\ion{He}{i}}	      &	    18	       &             0.198	       &    2.9$\pm$0.7        &       $-$               &       $-$               &      $-$             \\   
4069	&	[\ion{S}{ii}]	      &	    1F	       &             0.189	       &    0.64$\pm$0.3       &       4.6$\pm$1.4       &       1.9$\pm$0.5       &      $-$             \\   
4076	&	[\ion{S}{ii}]	      &	    1F	       &             0.187	       &    $-$                &       $-$               &       3.7$\pm$0.8       &      $-$             \\   
4102	&	{\ion{H}{i}}	      &	    H6	       &             0.182	       &    24.6$\pm$1.0       &       22.4$\pm$1.9      &       24.7$\pm$1.5      &      27.9$\pm$2.9    \\   
4144	&	{\ion{He}{i}}	      &	    53	       &             0.172	       &    1.1$\pm$0.2        &       $-$               &       $-$               &      $-$             \\   
4267	&	{\ion{C}{ii}}	      &	    6	       &             0.144	       &    0.33$\pm$0.09      &       $-$               &       $-$               &      $-$             \\   
4340	&	{\ion{H}{i}}	      &	    H$\gamma$  &             0.127	       &    44.9$\pm$1.0       &       44.5$\pm$2.1      &       46.6$\pm$1.2      &      46.5$\pm$3.0    \\   
4363	&	[\ion{O}{iii}]	      &	    2F	       &             0.121	       &    0.32$\pm$0.13      &       $-$               &       $-$               &      100.0$\pm$2.1   \\   
4388	&	{\ion{He}{i}}	      &	    51	       &             0.115	       &    0.48$\pm$0.11      &       $-$               &       $-$               &      $-$             \\   
4471	&	{\ion{He}{i}}	      &	    14	       &             0.096	       &    4.4$\pm$0.2        &       $-$               &       2.4$\pm$0.4       &      $-$             \\   
4639	&	{\ion{O}{ii}}	      &	    1	       &             0.055	       &    0.18$\pm$0.09      &       $-$               &       $-$               &      $-$             \\   
4649	&	{\ion{O}{ii}}	      &	    1	       &             0.052	       &    0.17$\pm$0.09      &       $-$               &       $-$               &      $-$             \\   
4651	&	{\ion{O}{ii}}	      &	    1	       &                	       &                       &                         &                         &                      \\   
4658	&	[\ion{Fe}{iii}]       &	    3F	       &             0.050	       &    0.32$\pm$0.09      &       $-$               &       $-$               &      $-$             \\   
4713	&	{\ion{He}{i}}	      &	    12	       &             0.036	       &    0.41$\pm$0.05      &       $-$               &       $-$               &      $-$             \\   
4861	&	{\ion{H}{i}}	      &	    H$\beta$	&            0.000	        &    100.0$\pm$2.0      &       100.0$\pm$2.2     &       100.0$\pm$2.1    &      $-$             \\   
4922	&	{\ion{He}{i}}	      &	    48	       &          $-$0.015	       &    1.2$\pm$0.12       &       $-$               &       0.24$\pm$0.12     &      $-$             \\   
4959	&	[\ion{O}{iii}]	      &	    1F	       &          $-$0.024	       &    25.6$\pm$0.6       &       $-$               &       13.9$\pm$0.5      &      3.8$\pm$0.4     \\   
5007	&	[\ion{O}{iii}]	      &	    1F	       &          $-$0.036	       &    79.1$\pm$1.7       &       $-$               &       41.9$\pm$0.9      &      11.3$\pm$0.7    \\   
5016	&	{\ion{He}{i}}	      &	    4   	   &          $-$0.038	       &    1.6$\pm$0.2        &       $-$               &       0.85$\pm$0.17     &      0.64$\pm$0.26   \\   
5198	&	[\ion{N}{i}]	      &	    1F	       &          $-$0.082	       &    0.26$\pm$0.03      &       1.8$\pm$0.4       &       1.9$\pm$0.1       &      1.2$\pm$0.3     \\   
5270	&	[\ion{Fe}{iii}]	      &	    1F	       &          $-$0.100	       &    0.14$\pm$0.03      &       $-$               &       0.68$\pm$0.10     &      $-$             \\   
5518	&	[\ion{Cl}{iii}]       &	    1F	       &          $-$0.154	       &    0.39$\pm$0.05      &       $-$               &       0.40$\pm$0.12     &      $-$             \\   
5538	&	[\ion{Cl}{iii}]       &	    1F	       &          $-$0.158	       &    0.28$\pm$0.04      &       $-$               &       0.24$\pm$0.09     &      $-$             \\   
5666	&	{\ion{N}{ii}}	      &	    3	       &          $-$0.180        &    0.07$\pm$0.03      &       $-$               &       $-$                &      $-$             \\   
5679	&	{\ion{N}{ii}}	      &	    3	       &          $-$0.182        &    0.07$\pm$0.03      &       $-$               &       $-$                &      $-$             \\   
5755	&	[\ion{N}{ii}]	      &	    3F	       &          $-$0.194	       &    0.48$\pm$0.05      &       0.52$\pm$0.19     &       1.14$\pm$0.08     &      1.4$\pm$0.2     \\   
5876	&	{\ion{He}{i}}	      &	    11	       &          $-$0.215	       &    13.7$\pm$0.3       &       2.0$\pm$0.4       &       7.18$\pm$0.22     &      4.5$\pm$0.2     \\   
6300	&	[\ion{O}{i}]	      &   	1F	       &          $-$0.282	       &    0.31$\pm$0.04      &       $-$               &       0.97$\pm$0.15     &      1.7$\pm$0.2     \\   
6312	&	[\ion{S}{iii}]	      &   	3F	       &          $-$0.283	       &    0.74$\pm$0.08      &       $-$               &       1.0$\pm$0.2       &      0.53$\pm$0.12   \\   
6347	&	{\ion{Si}{ii}}	      &	    4	       &          $-$0.289        &    0.13$\pm$0.05      &       $-$               &       0.39$\pm$0.12      &      0.34$\pm$0.16   \\   
6363	&	[{\ion{O}{i}}]	      &	    1F	       &          $-$0.291        &   $-$                 &       $-$               &       0.35$\pm$0.10      &      0.43$\pm$0.15   \\   
6371	&	{\ion{Si}{ii}}	      &	    2	       &          $-$0.292        &   $-$                 &       $-$               &       0.23$\pm$0.08      &      $-$             \\   
6548    &   [\ion{N}{ii}]          &    1F         &          $-$0.318        &    24.9$\pm$0.7       &       37.7$\pm$2.3      &       41.9$\pm$1.2       &      60.4$\pm$2.7    \\   
6563    &   {\ion{H}{i}}          &     H$\alpha$  &          $-$0.320        &    281.7$\pm$7.7      &       269.0$\pm$16.2    &       283.2$\pm$8.0      &      305.1$\pm$13.8  \\   
6583    &   [\ion{N}{ii}]         &     IF         &          $-$0.323        &    76.0$\pm$2.1       &       102.2$\pm$6.2     &       129.5$\pm$0.8      &      181.5$\pm$8.2   \\   
6678    &   {\ion{He}{i}}         &     46         &          $-$0.336        &    3.3$\pm$1.0        &       0.64$\pm$0.13     &       1.9$\pm$0.08       &      1.2$\pm$0.1     \\   
6716    &   [\ion{S}{ii}]         &     2F         &          $-$0.342        &    13.5$\pm$0.4       &       41.6$\pm$2.7      &       15.1$\pm$0.4       &      31.9$\pm$1.5    \\   
6731    &   [\ion{S}{ii}]         &     2F         &          $-$0.344        &    11.2$\pm$0.3       &       29.6$\pm$1.9      &       18.3$\pm$0.5       &      32.6$\pm$1.6    \\   
7065	&	{\ion{He}{i}}	      &	    10     	   &          $-$0.387	       &    2.1$\pm$0.08       &       1.2$\pm$0.2       &       2.8$\pm$0.1       &      0.95$\pm$0.09   \\   
7136    &   [\ion{Ar}{iii}]       &     1F         &          $-$0.396        &    8.1$\pm$0.2        &       0.81$\pm$0.18     &       6.7$\pm$0.2        &      2.9$\pm$0.2     \\   
7318	&	[\ion{O}{ii}]	      &	    2F	       &          $-$0.418	       &    0.65$\pm$0.13      &       $-$               &       3.5$\pm$0.2       &      2.2$\pm$0.3     \\   
7320	&	[{\ion{O}{ii}}]	      &	    2F	       &                          &                       &                         &                          &                      \\   
7330	&	[\ion{O}{ii}]	      &	    2F	       &          $-$0.420	       &    0.59$\pm$0.12      &       $-$               &       2.8$\pm$0.2       &      1.8$\pm$0.2     \\   
7331	&	[\ion{O}{ii}]	      &	    2F	       &                 	       &                       &                         &                         &                      \\   
7751    &   [\ion{Ar}{iii}]       &     2F         &         $-$0.467         &    1.4$\pm$0.2        &       $-$               &       1.3$\pm$0.2        &      1.1$\pm$0.2     \\   

\hline
$c$(H$\beta$)     &              &                 &                          &  1.83$\pm$0.03      &     1.49$\pm$0.08        &    2.46$\pm$0.03        &   2.56$\pm$0.05        \\
$F$(H$\beta$)$^{\rm a}$     &              &                 &                          &    3.98$\pm$0.08   &    1.02$\pm$0.02  &     2.28$\pm$0.05  &  0.94$\pm$0.02  \\
\hline
\multicolumn{8}{l}{$^{\rm a}$ Flux uncorrected for reddening in units of 10$^{-14}$ erg cm$^{-2}$ s$^{-1}$.}
\end{longtable}

\section{Additional results for N/H and N/O gradients} 
\begin{figure} 
\begin{center}
    \includegraphics[width=0.7\textwidth, trim=30 0 30 0,  clip=yes]{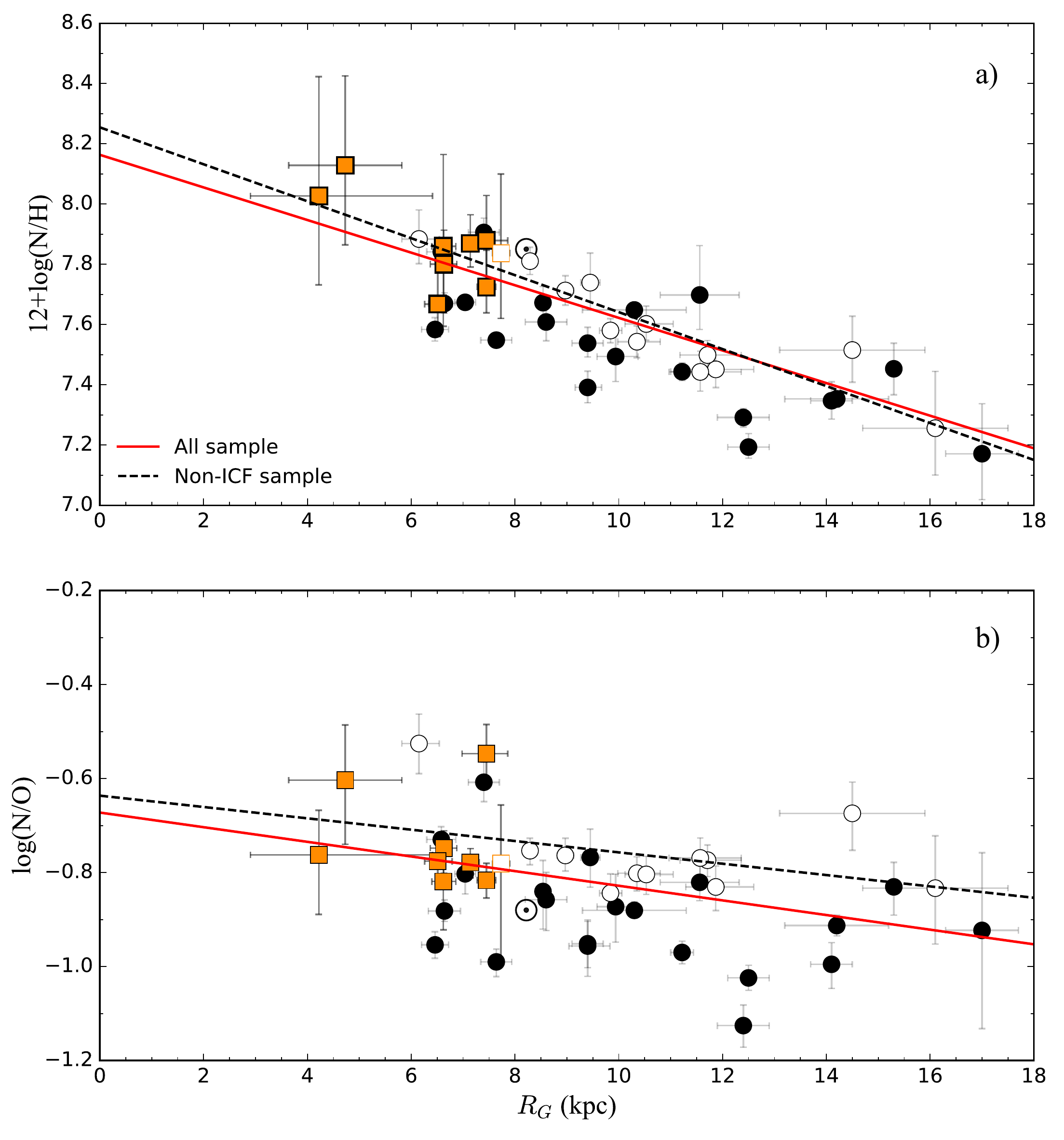}
    \caption{Same as Fig.~\ref{fig:N-gradient} and~\ref{fig:NO-gradient} but using the ICF of \citet{Peimbert:1969} for determining the N/H and N/O ratios of the objects represented with orange or black filled symbols, respectively. Comparing with Figs.~\ref{fig:N-gradient} and ~\ref{fig:NO-gradient}, we can see that the match between the filled symbols and the empty ones (which represent objects whose N/H ratio has been derived without an ICF) is not as good as in Figs.~\ref{fig:N-gradient} and ~\ref{fig:NO-gradient}. Also, the dispersion of the points around the fits representing the N/H and N/O gradients is larger in this case.}
    \label{fig:N-gradient-PC69}
   \end{center}
    \end{figure}

In this appendix, we include two figures representing the same as figs.~\ref{fig:N-gradient} and 
\ref{fig:NO-gradient}, but with the N abundances calculated using the ICF scheme of \citet{Peimbert:1969}. The panels (a) and (b) of Fig. \ref{fig:N-gradient-PC69} show also the N/H and N/O gradients as a function of Galactocentric distances. The symbols are the same as in figs. \ref{fig:N-gradient} and \ref{fig:NO-gradient}. We obtain linear gradients of 12+log(N/H) = (8.14$\pm$0.07)  $-$ (0.055$\pm$0.007) $\times$ $R_G$ and log(N/O) = ($-$0.67$\pm$0.05) $-$ (0.016$\pm$0.005) $\times$ $R_G$, with dispersion around the fit of 0.12 dex and 0.11 dex, respectively. A comparison and discussion of these results is presented in \S~\ref{sec:nitrogen}.

\bsp	
\label{lastpage}
\end{document}